\UseRawInputEncoding
\pdfoutput=1
\documentclass[acmsmall]{acmart}
\AtBeginDocument{%
  }

\setcopyright{none}
\renewcommand\footnotetextcopyrightpermission[1]{}
\usepackage{balance}
\usepackage{tcolorbox}
\usepackage{bbm}
\usepackage{enumerate}
\usepackage{algorithm}
\usepackage{algorithmicx}
\usepackage[noend]{algpseudocode}
\usepackage{multirow}
\usepackage{pifont}
\usepackage{graphicx}
\graphicspath{{figures/}}
\usepackage{subcaption}
\usepackage{hyperref}
\usepackage{listings}
\usepackage{xcolor}
\usepackage{enumitem}

\definecolor{mygreen}{rgb}{0,0.6,0}  
\definecolor{mygray}{rgb}{0.5,0.5,0.5}  
\definecolor{mymauve}{rgb}{0.58,0,0.82}  

\definecolor{codegreen}{rgb}{0,0.6,0}
\definecolor{codegray}{rgb}{0.5,0.5,0.5}
\definecolor{codepurple}{rgb}{0.58,0,0.82}
\definecolor{backcolour}{rgb}{0.95,0.95,0.92}

\lstdefinestyle{mystyle}{
    backgroundcolor=\color{backcolour},   
    commentstyle=\color{codegreen},
    keywordstyle=\color{magenta},
    numberstyle=\tiny\color{codegray},
    stringstyle=\color{codepurple},
    basicstyle=\ttfamily\footnotesize,
    breakatwhitespace=false,         
    breaklines=true,                 
    captionpos=b,                    
    keepspaces=true,                 
    numbers=left,                    
    numbersep=5pt,                  
    showspaces=false,                
    showstringspaces=false,
    showtabs=false,                  
    tabsize=2
}

\graphicspath{{figures/}}

\newcommand{\figref}[1]{Figure~\ref{#1}}
\newcommand{\tabref}[1]{Table~\ref{#1}}

\usepackage{hyperref}
\hypersetup{
  colorlinks   = true,    
  urlcolor     = blue,    
  linkcolor    = blue,    
  citecolor    = blue      
}

\usepackage{xspace}
\newcommand{\name}{\texttt{LEADE}\xspace}

\begin{document}

\title{LMM-enhanced Safety-Critical Scenario Generation for Autonomous Driving System Testing From Non-Accident Traffic Videos}

\author{Haoxiang Tian}
\affiliation{
  \institution{Institute of Software Chinese Academy of Sciences, University of Chinese Academy of Sciences}
  \country{China}
}
\email{tianhaoxiang20@otcaix.iscas.ac.cn}

\author{Xingshuo Han}
\affiliation{
  \institution{Nanyang Technological University}
  \country{Singapore}
}
\email{xingshuo001@e.ntu.edu.sg}

\author{Yuan Zhou}
\affiliation{
  \institution{Zhejiang Sci-Tech University}
  \country{China}
}
\email{yuanzhou@zstu.edu.cn}


\author{Guoquan Wu}
\affiliation{
  \institution{Institute of Software Chinese Academy of Sciences}
  \country{China}
}
\email{gqwu@otcaix.iscas.ac.cn}

\author{An Guo}
\affiliation{
  \institution{Nanjing University}
  \country{China}
}
\email{guoan218@smail.nju.edu.cn}

\author{Mingfei Cheng}
\affiliation{
  \institution{Singapore Management University}
  \country{China}
}
\email{snowbirds.mf@gmail.com}

\author{Shuo Li}
\author{Jun Wei}
\affiliation{
  \institution{Institute of Software Chinese Academy of Sciences}
  \country{China}
}
\email{{lishuo19, wj}@otcaix.iscas.ac.cn}

\author{Tianwei Zhang}
\affiliation{
  \institution{Nanyang Technological University}
  \country{Singapore}
}
\email{tianwei.zhang@ntu.edu.sg}

\renewcommand{\shortauthors}{Haoxiang Tian et al.}

\begin{abstract}
Safety testing serves as the fundamental pillar for the development of autonomous driving systems (ADSs). To ensure the safety of ADSs, it is paramount to generate a diverse range of safety-critical test scenarios. 
While existing ADS practitioners primarily focus on reproducing real-world traffic accidents in simulation environments to create test scenarios, it's essential to highlight that many of these accidents do not directly  result in safety violations for ADSs due to the differences between human driving and autonomous driving.
More importantly, we observe that some accident-free real-world scenarios can not only lead to misbehaviors in ADSs but also be leveraged for the generation of ADS violations during simulation testing. Therefore, it is of significant importance to discover safety violations of ADSs from routine traffic scenarios (i.e., non-crash scenarios) to ensure the safety of Autonomous Vehicles (AVs).

We introduce \name, a novel methodology to achieve the above goal. It automatically generates abstract and concrete scenarios from real-traffic videos. Then it optimizes these scenarios to search for safety violations of the ADS in semantically consistent scenarios where human-driving worked safely. Specifically, \name enhances the ability of Large Multimodal Models (LMMs) to accurately construct abstract scenarios from traffic videos and generate concrete scenarios by multi-modal few-shot Chain of Thought (CoT). Based on them, \name assesses and increases the behavior differences between the ego vehicle (i.e., the vehicle connected to the ADS under test) and human-driving in semantic equivalent scenarios (here equivalent semantics means that each participant in test scenarios has the same abstract behaviors as those observed in the original real traffic scenarios). We implement and evaluate \name on the industrial-grade Level-4 ADS, Apollo. The experimental results demonstrate that compared with state-of-the-art ADS scenario generation approaches, \name can accurately generate test scenarios from traffic videos, and effectively discover more types of safety violations of Apollo in test scenarios with the same semantics of accident-free traffic scenarios.
\end{abstract}



\maketitle

\section{Introduction}
\label{sec:introduction}
Autonomous Driving System (ADS) testing is one of the most important tasks for the safety assurance of Autonomous Vehicle (AV) software \cite{feng2021survey}. It is essential to perform thorough testing on ADSs before deploying them in the real world \cite{barr2014oracle}. This process requires large amounts of diverse and comprehensive traffic scenarios. However, physical testing on public roads takes huge costs and increases the risk of accidents. Alternatively, simulation testing can create massive scenarios with extremely low costs. According to the Autonomous Driving Simulation Industry Chain Report \cite{chain_report}, over 80\% of ADS testing is completed through simulation platforms.




The main step in simulation testing is to generate safety-critical test scenarios, from which we can find various safety violations of ADSs \cite{singh2021autonomous, chen2023end}. Several companies \cite{nvidia_strive, waymo} and academic researchers \cite{10.1145/3338906.3338942, zhang2023building, guo2024sovar} have reproduced real-world accidents in the simulator to construct the test scenarios. In these scenarios, the tester drives the ego vehicle (the vehicle controlled by the ADS under test) following the route of a vehicle involved in the collision, and defines the trajectories of other participants as those of other vehicles and pedestrians in the accident. However, there are two shortcomings in this strategy. First, in real-traffic accidents, more than 55\% crashes \cite{nhtsa_files} are not effectively linked to the safety violations of ADSs, such as the ones caused by alcohol, cell phone use, fatigue, stress, drivers' physical and internal illness, illegal maneuvers(e.g., retrograding) \cite{singh2015critical, almaskati2023autonomous}, which will not occur to ADSs or are not the responsibility of ADSs. 
Second, due to the inherent differences in decision-making and operations between human drivers and ADSs, some safe human-driving traffic scenarios can actually cause safety violations of ADSs.

\begin{figure}[t]
  \centering
  \includegraphics[scale=0.5]{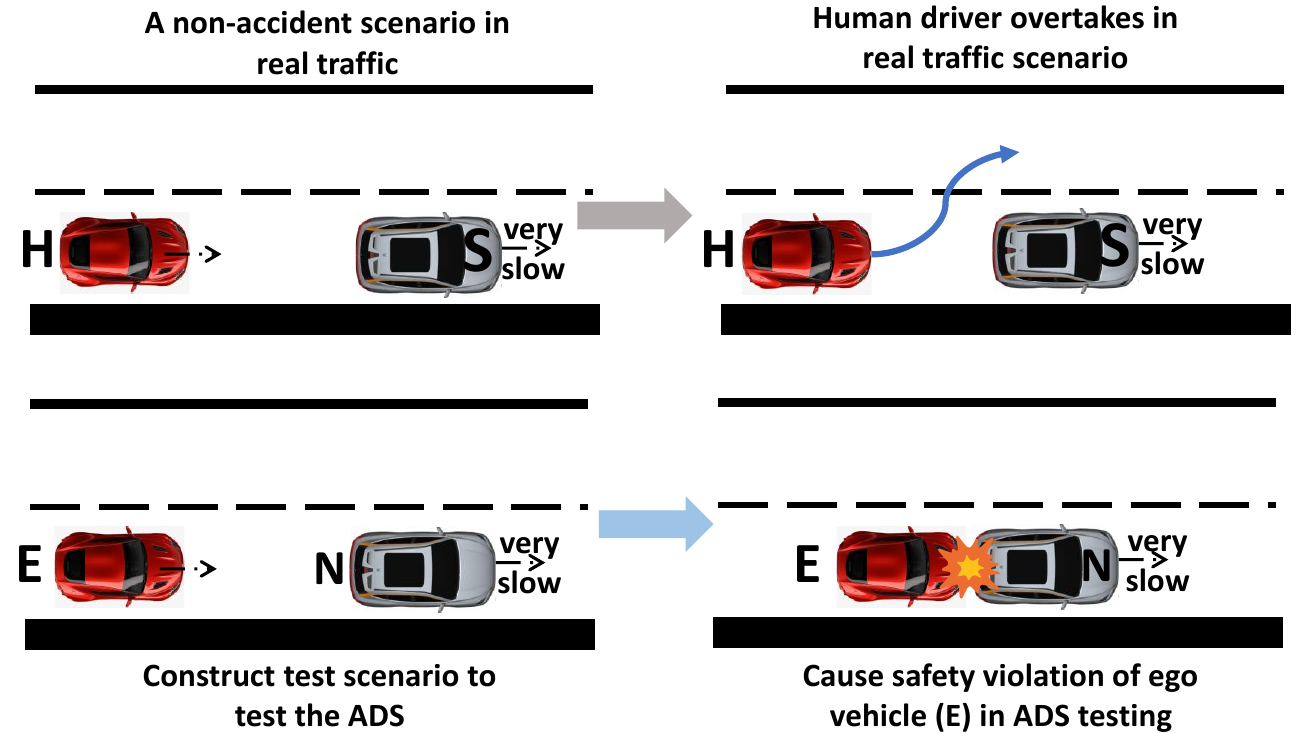}
  \caption{An example of a real-traffic scenario without accident that can cause safety violations of ADSs.}
  \label{motivation}
  \vspace{-10px}
  \end{figure}

Figure~\ref{motivation} shows one example. In the human-driving case, a vehicle $S$ is driving at a very slow speed. The human driver of the following vehicle $H$ can safely overtakes $S$ and no accident occurs. We then apply the same scenario to the simulation testing of the Apollo ADS \cite{apolloauto}, where the NPC vehicle $N$ is moving slowly. Initially, the ADS of the following ego vehicle $E$ regards $N$ as stationary due to its very slow speed, and generates the future path for overtaking $N$. However, during runtime, the motion of $N$ makes its distance to $E$ vary continuously, which mislead $E$ not to perform overtake and get stuck for a long time, finally colliding with $N$. 
In summary, assessing the safety of ADSs through real-world traffic accidents is insufficient. It is equally important to conduct comprehensive safety evaluations of ADSs in accident-free human-driving scenarios, an area that is rarely investigated.



Motivated by the above observations, this paper proposes \name,
a novel approach that discovers safety violations of ADSs from real accident-free human-driving traffic scenarios. Our source of traffic scenarios is the automotive videos, which are commonly available and completely record the static and dynamic surrounding objectives \cite{ghahremannezhad2022traffic}. \textbf{The goal of \name} is to discover safety violations of the ADS in test scenarios while keeping the equivalent semantics as the original accident-free traffic scenarios. Here, the equivalent semantics means that in the test scenario, each participant has the same abstract behaviors (e.g., lane changing) as those observed in the original real traffic scenario. In the following, we call such test scenarios as the \textit{semantic equivalent scenarios} for simplicity.
Specifically, to maintain the semantic equivalence during search, \name first constructs abstract scenarios from traffic videos, and then converts them into executable concrete scenarios. Based on these abstract and concrete scenarios, \name searches for safety-critical semantic equivalent scenarios, which can reflect the differences between the behaviors of the ego vehicle and human drivers.
However, there exist the following challenges to achieve the above process:

\begin{itemize}[leftmargin=*]
\item \textbf{Challenge 1:} How to design and implement a lightweight method to accurately understand the real-traffic videos, and generate the abstract and executable concrete scenarios correctly?
\item \textbf{Challenge 2.} How to evaluate the behavior differences between the ego vehicle in a test scenario and the human driver in the original scenario?
\item \textbf{Challenge 3.} How to search for safety-critical semantic equivalent scenarios, where ego vehicle's behaviors in test scenarios are different from human-driving behaviors in traffic scenarios?
\end{itemize}

\name consists of two innovative techniques to address the above challenges. \textbf{For challenge 1}, we introduce a method for \textbf{traffic video-based scenario understanding and generation}. Inspired by the great contextual text-image understanding and rule reasoning capability of Large Multimodal Models (LMMs) \cite{gpt4-corner, gpt4-corner2}, \name first utilizes an LMM to understand scenarios of real-traffic videos and construct abstract scenarios from them. As currently LMMs do not support the input format of videos, \name utilizes optical flow analysis and state interpolation to enhance the LMM's ability in understanding, extracting and constructing the scenario semantics. Then instead of defining trajectory generation rules for each type of behavior, \name utilizes multi-modal few-shot Chain-of-Thought (CoT) to guide the LMM to generate executable concrete scenario programs from abstract scenarios.

\textbf{For challenge 2}, we introduce a \textbf{dual-layer search for safety-critical semantic equivalent test scenarios }. It refines the search space to maximize the differences between the behaviors of the ego vehicle in the test scenario and human-driving in the original scenario, and verifies the universality of ADS's safety violations while keeping the equivalent semantics of generated test scenarios during the search. The behavior difference is more robust to reflect the discrepancies in driving intentions and decision-making than the trajectory differences. In our solution, the outer-layer assesses and increases the behavior differences, while the inner-layer explores the variations of participants' trajectories with equivalent semantics of corresponding traffic videos, verifying the universality of the exposed safety violations.


To evaluate the effectiveness and efficiency of \name, we conduct evaluations on Baidu Apollo \cite{apolloauto} in the SVL simulator \cite{sora-svl}. Experimental results show that \name can 
correctly generate abstract scenarios and executable concrete scenarios from real-traffic videos that contain various types of roads and behaviors. Furthermore, it can efficiently generate safety-critical semantic equivalent scenarios. Consequently, \name exposes 10 distinct types of safety violations in Apollo, 7 of which are new and never discovered by existing state-of-the-art solutions.

The contributions of this paper are as follows: 
\begin{itemize}[leftmargin=*]
    \item We introduce \name, a novel approach to automatically identify the safety violations of ADSs from accident-free real-traffic videos. 
    \item \name enhances LMM's ability to accurately construct abstract scenarios from traffic videos, and generate the corresponding concrete scenarios with multi-modal few-shot CoT prompts. Based on them, \name assesses and increases the behavior differences between ego vehicle and human-driving in semantic equivalent scenarios. It utilizes a dual-layer optimization search fo discover safety violations of the ADS in semantic equivalent scenarios. 
    \item We test \name on an industrial L4 ADS, Apollo \cite{apolloauto}. The experimental results demonstrate the effectiveness of \name. Compared with state-of-the-art approaches, \name can generate abstract and concrete scenarios from automotive videos more accurately, and discover more types of safety violations of Apollo.
\end{itemize}

\section{Background and Related Work}




\subsection{ADS Safety Testing}

Given the significant impact of the ADS, it is important to comprehensively test its safety before deployed on real roads. The key idea is to generate diverse safety-critical \textit{scenarios}, and measure the behaviors of the vehicle in simulation when encountering these scenarios. This provides valuable feedback to improve the internal designs and algorithms of the ADS.

\subsubsection{Semantics of Test Scenarios}
The semantics of ADS test scenarios refer to the formal definition and interpretation of scenario elements that describe real-world driving situations \cite{hungar2018scenario, zipfl2023comprehensive}. It is essential to consider factors such as road types, traffic participants and environmental conditions. Testing ADSs with simulation emphasizes the dynamic interactions between the ego vehicle with moving objects in scenarios. Therefore, in this work, we focus on the traffic participants (e.g., NPC vehicles, pedestrians) for ADS test scenario generation. The semantics of test scenarios contain the high-level abstraction of these participants: road type, types and behaviors of NPC vehicles and pedestrians, relative positions of NPC vehicles and pedestrians to the ego vehicle.

\subsubsection{Descriptions of Test Scenarios}
Testing scenarios are normally described by \textit{scenario programs}, which are executable in the simulator.
Existing simulation platforms provide low-level APIs to construct driving scenarios \cite{lou2022testing}, scenario programs based on these APIs require loads of codes and cumbersome syntax. For example, OpenScenario \cite{jullien2009openscenario} is a popular framework for describing the dynamic scenarios in test drive simulations. It takes 258 lines of code to construct a simple driving scenario of single vehicle's lane cutting. The scenarios described by low-level APIs are not conducive to high-level comprehension, understanding and reasoning of ADS testing.


  
\subsubsection{Generation of Test Scenarios}
The effectiveness of the testing highly depends on the quality of the generated scenarios. Due to the huge input space and functional complexity of ADSs \cite{shin2018test, nejati2019testing, arrieta2017search}, it becomes infeasible for conventional software testing approaches to capture various rare events from complex driving situations~\cite{guo2024semantic}. How to generate diverse safety-critical scenarios to fully test ADSs has received significant attention in recent years \cite{10.1145/2884781.2884880, hekmatnejad2020search, cao2021work}. 

Some works reproduce traffic accidents from crash databases as the test scenarios~\cite{10.1145/3338906.3338942, zhang2023building, guo2024sovar, tang2024legend}. Specifically, AC3R \cite{10.1145/3338906.3338942} uses domain-specific ontology and NLP to extract information from police crash reports and reconstruct corner cases of crash accidents. SoVAR \cite{guo2024sovar} utilizes the LLM to extract comprehensive key information from police crash reports, and formulates them to generate corresponding test scenarios. However, both methods require crash reports that follow rigid narrative standards and record detailed text information of accidents, which is not applicable to free-form scenarios, e.g., automotive videos from real-life vehicles, to executable concrete scenarios. Zhang et al. \cite{zhang2023building} train a panoptic segmentation model, M-CPS, to extract effective information from the accident videos and recover traffic participants in the simulator. However, it is time-consuming to train and optimize the extraction model. Furthermore, the accuracy and correctness of M-CPS in key information extraction is not high. More importantly, a common limitation of these methods is that they are only applicable for traffic accidents, but not feasible to generate safety-critical scenarios from massive regular traffic situations. Traffic crashes are rare in real life and most of them cannot create effective challenges to ADSs \cite{singh2015critical, almaskati2023autonomous}. As a result, the test scenarios that reproduce traffic accidents can not fully test industrial ADSs and find their safety violations comprehensively. Different from these methods, \name generates executable test scenarios from real-traffic scenarios and searches for safety violations of the ADS in semantic equivalent scenarios.

\section{Methodology} 

\begin{figure}[t]
  \centering
  \includegraphics[scale=0.25]{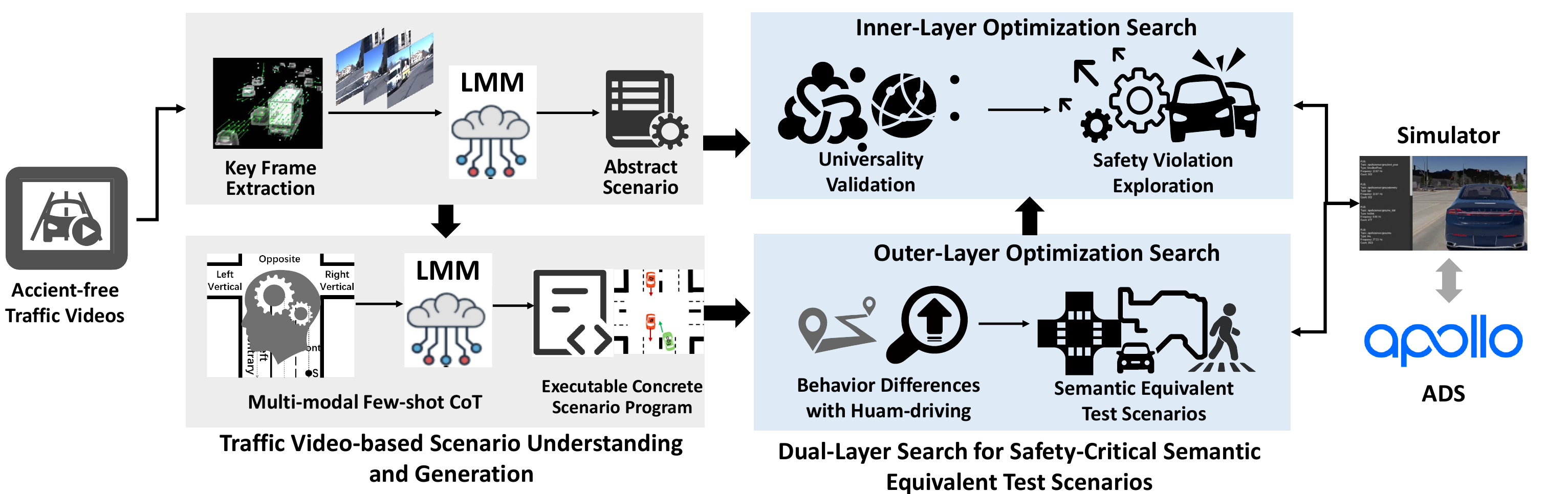}
  \caption{The overview of \name.}
  \label{overview}
  \vspace{-10px}
  \end{figure} 

\figref{overview} shows the overview of \name, which consists of two critical parts:

\noindent \textbf{(1) Traffic Video-based Scenario Understanding and Generation} (Section \ref{sec:method-understand}). The input of \name is automotive videos. It first utilizes optical flow analysis and state interpolation to understand the scenario semantics from traffic videos with LMMs. Based on these, it constructs abstract scenarios and generates concrete scenarios conforming to the semantics of these traffic videos. Instead of defining a large set of rules for concrete scenario generation, \name utilizes multi-modal few-shot CoT to guide LMMs to generate executable concrete scenario programs.

\noindent \textbf{(2) Dual-Layer Search for Safety-Critical Semantic Equivalent Test Scenarios} (Section \ref{sec:method-search}). Based on the generated abstract and concrete scenarios, \name designs a dual-layer optimization search to find safety violations of the ADS in semantic equivalent scenarios. Specifically, the outer-layer optimization leverages the concrete scenarios to explore the differences between the behaviors of the ego vehicle and human-driving for the same driving task in semantic equivalent scenarios. The inner-layer optimization search is to verify whether the discovered safety violation of the ADS is universal in the traffic situations of the abstract scenario.

\subsection{Traffic Video-based Scenario Understanding and Generation}
\label{sec:method-understand}
Traffic videos provide rich and complete information of real traffic scenarios, and have become one of the most common and accessible data sources of real-world traffic \cite{ghahremannezhad2022traffic}. To generate test scenarios from real traffic, \name first understands static and dynamic semantics of traffic scenarios from a collection of automotive videos. 
Past works have demonstrated that LMMs, represented by GPT-4V \cite{gpt4v1, gpt4v2}, have the strong capabilities in accurately recognizing and understanding the elements in driving scenarios \cite{gpt4-corner, gpt4-corner2}. However, state-of-the-art LMMs do not support the analysis of videos, and their input size is limited. To address this gap, \name designs a motion change-based key frame extraction technique, which uses optical flow analysis and state interpolation to identify motion changes of participants in the scenario as key frames. Then it leverages the LMM to understand and extract the information of scenario dynamics from the identified key frame sequence of traffic videos.

Based on the scenario semantics of traffic videos, \name generates the corresponding test scenarios in the simulation environment for ADS testing. To improve the generalizability of the generated test scenarios, \name first constructs abstract scenarios, which serves as a high-level representation. They are typically characterized by road topology, traffic lights, and general behaviors of dynamic participants (e.g., a pedestrian crossing a road in front of the ego vehicle, or a following car maintaining a certain speed). Then it generates executable concrete scenarios based on the abstract scenarios.   

Existing methods of generating concrete scenarios from abstract scenarios \cite{tian2022generating, fremont2019scenic, guo2024sovar} require defining a large set of rules and constraints for each participant behavior, which is not flexible or conducive to extension. To address the limitation, \name utilizes multi-modal few-shot Chain-of-Thought (CoT) to guide the LMM for the generation, by leveraging LMM's contextual text-image learning and rule reasoning capability \cite{romera2024mathematical}. Below we give details of the above process. 

\subsubsection{Motion Change-Based Key Frame Extraction.}
\label{sec:motion-change-extraction}
Previous approaches to key frame extraction, such as uniform sampling \cite{ishak2014developing} and attention-based methods \cite{shih2013novel, ejaz2013efficient}, achieve varying degrees of success in identifying important frames. However, as real-world traffic environments are complex with various participant behaviors, these methods struggle to balance between computational efficiency and the ability to capture non-uniform frame transitions. 

\name designs a new approach that dynamically identifies key frames based on the motion changes of vehicles and pedestrians, such as speed shifts and direction changes. Specifically, it leverages optical flow analysis to obtain the motion states of vehicles and pedestrians over time in traffic videos. Then for each frame $f$, based on the motion states of its former and next frames, it employs linear interpolation to estimate the intermediate state between them. If the participant's motion state does not change between two frames, this intermediate state should be equal or close to the participant's actual motion state at the frame $f$. For the frames where the motion states of participants change, they are sequenced as key frames of the traffic scenario.

First, \name segments the input traffic video according to the duration to generate the sequence of scenario frames. It leverages optical flow analysis \cite{wang2020abnormal} to separate the moving objects from the background and generate optical flow field vector for the moving objects. To reduce the computation time cost and improve the accuracy of obtaining dynamic objects information, \name converts the sequence of the frames into the gray-scale format, and generates the optical flow field vector that contains information about the motion states of objects in consecutive frames. The optical flow field vector of a scenario is formed by the displacement of the corresponding pixels between consecutive frames caused by the motion objects. 
For the traffic video segmented as $N$ frames, the optical field vector consists of $N$ motion state vectors. At frame $f$, the motion state vector is represented as:
\begin{equation}
    \Vec{S_f} = \left[ \begin{array}{cccc}
  \mathbb{X}_1^f & \mathbb{Y}_1^f & \mathbb{VX}_1^f & \mathbb{VY}_1^f \\  
  \mathbb{X}_2^f & \mathbb{Y}_2^f & \mathbb{VX}_2^f & \mathbb{VY}_2^f \\
  ...... & ...... & ...... & ...... \\
  \mathbb{X}_K^f & \mathbb{Y}_K^f & \mathbb{VX}_K^f & \mathbb{VY}_K^f
\end{array} \right]
\end{equation}
where $K$ represents the number of dynamic objects at the frame. $\mathbb{X}_i$ and $\mathbb{Y}_i$ are the coordinates with respect to the camera origin of the pixels in the $i$-th dynamic object, and $\mathbb{VX}_i$ and $\mathbb{VY}_i$ are the horizontal and vertical velocity components of the object. Here \name adopts the Lucas Kanade algorithm \cite{plyer2016massively} in the optical flow analysis. 

Based on the optical flow vectors of the traffic scenario, \name uses linear interpolation \cite{sarker2020screw} to identify the significant changes of motion states between frames. 
For the motion state vector $\Vec{S_f}$ at frame $f$, \name selects the vectors of its former frame $\Vec{S_{f-1}}$ and next frame $\Vec{S_{f+1}}$. Then it uses linear interpolation to compute the intermediate vector between $\Vec{S_{f-1}}$ and $\Vec{S_{f+1}}$ in a linear progression. The intermediate state $\Vec{P'_{f}}$ is computed as:
\begin{equation}
    \Vec{P'_f} = (1 - \alpha) \Vec{S_{f-1}} + \alpha \Vec{S_{f+1}}, \alpha \in [0, 1]
\end{equation}
where $\alpha$ represents the interpolation parameter. \name measures the difference between the actual motion state vector $\Vec{S_f}$ and the intermediate vector $\Vec{P'_f}$ to identify the motion changes at frame $f$. The non-linear changes of dynamic objects correspond to key events (e.g., sudden stops, lane changes, accelerations) in their movements. Therefore, for all frames of the scenario, the non-linear changes are identified to extract key frames, which are added to the key frame sequence if the interpolated state deviates significantly from the real state. Formally, the difference between the actual motion state vector $\Vec{S_f}$ and the intermediate vector $\Vec{P'_f}$ is calculated as follows:
\begin{equation}
    \Delta M(\Vec{S_f}, \Vec{P'_f}) = \sum_{x,y} \left\| L_{x,y}(\Vec{S_f}, \Vec{P'_f}), O_{x,y}(\Vec{S_f}, \Vec{P'_f}) \right\|
\end{equation}
where $L_x$ and $L_y$ are the differences of motion state vectors in the horizontal and vertical positions respectively; $O_x$ and $O_y$ are the differences of motion state vectors in the horizontal and vertical velocities respectively. $\Vert \cdot \Vert$ denotes a L2-norm to quantify the difference between motion states. A significant deviation (i.e., above a pre-defined threshold $\tau$) between the actual motion state vector and the expected linear motion state vector. 

\subsubsection{Abstract Scenario Construction}

For the extracted key frames of a traffic video, \name utilizes the LMM to understand and recognize the static and dynamic elements in the scenario, including road type and structure, types and behaviors of dynamic participants (e.g., vehicles, pedestrians) in the video. As the performance of LMMs heavily depend on the quality of prompts \cite{gpt_quality}, we design linguistic patterns to generate the following input prompt for the LMM, promoting it to understand the static and dynamic elements in the traffic scenario:
\begin{center}
\begin{tcolorbox}[colback=gray!10,
                  colframe=black,
                  width=\linewidth,
                  arc=1mm, auto outer arc,
                  boxrule=0.5pt,
                 ]
You are an autonomous driving expert who specializes in identifying dynamic objects in traffic scenarios. I will show you a series of traffic pictures taken by the camera of the vehicle you are driving. These pictures are from the same one scenario. Please use concise and structured language to describe the following objects in the scenario: road types, behaviors of the vehicle you are driving, behaviors and positions of traffic participants (all vehicles and pedestrians, other signals and obstacles within the visible range).
\end{tcolorbox}
\end{center}
Based on the extracted information of scenario elements, \name constructs the abstract scenarios by leveraging NLP to parse them into semi-structured description contents. Specifically, it uses StandfordNLP \cite{manning2014stanford} to perform part-of-speech tagging and dependency analysis on the corresponding extracted results, which converts the extracted results into semi-structured descriptions including the following attributes: road type, vehicle type, vehicle role, behaviors, relative initial positions. Table~\ref{abs} gives an example of an abstract scenario.


\begin{table}[h]\small
\vspace{-5px}
\caption{An example of an abstract scenario}
\vspace{-5px}
\renewcommand\arraystretch{1.3}
\begin{tabular}{|c|c|c|c|c|}
\hline
\textbf{road type}            & \textbf{vehicle role} & \textbf{type} & \textbf{behaviors}      & \textbf{\begin{tabular}[c]{@{}c@{}}relative position\end{tabular}} \\ \hline
\multirow{3}{*}{intersection} & ego                   & car           & change lane, turn right & —                                                                    \\ \cline{2-5} 
                              & NPC                   & truck         & follow lane, cross      & left front                                                           \\ \cline{2-5} 
                              & Pedestrian            & —             & stand, cross            & right vertical                                                       \\ \hline
\end{tabular}
\label{abs}
\vspace{-5px}
\end{table}

\subsubsection{Concrete Scenario Program Generation}
\label{concrete}
Based on each abstract scenario, 
\name designs a lightweight method to generate executable test scenario programs with the following two steps. First, We study the practice of executable test scenario programs (an example is shown in List~\ref{program}), and decomposes the concrete scenario program into five components: map statement, ego driving task statement, NPC vehicles' trajectories definition, pedestrians' trajectories definition, scenario assertion statement. Their descriptions are as follows: 
 
\begin{enumerate}[label=(\textbf{\alph*}), leftmargin=*]

\item The map statement specifies the map used in the simulator. 

\item The ego driving task statement defines the type, initial position and destination of ego vehicle. The initial position and destination make the ego vehicle plan a route that corresponds to its behavior and the road type specified in abstract scenario. 

\item The NPC vehicles' trajectories definition defines the type and waypoints of NPC vehicles in the scenario, which make them perform the corresponding behavior specified in the abstract scenario. 

\item The pedestrians' trajectories definition defines the waypoints of pedestrians in the scenario, which make them perform the corresponding behavior specified in the abstract scenario. 

\item The scenario assertion statement is to verify whether the ADS behavior aligns with safety requirements. Furthermore, according to the assertion statement, the values of the required variables for assertion are collected during the execution of the scenario.

\end{enumerate}

Based on the above five components, the concrete scenario program generation is divided into five stages: map selection, ego driving task designation, NPC vehicle trajectory generation, pedestrian trajectory generation, and assertion definition.


Second, \name uses the multi-modal few-shot Chain-of-Thought (CoT) prompting to generate the executable concrete scenarios programs for the given abstract scenario. 
This is inspired by the practice that the LMM performs well in identifying behaviors from trajectories in traffic scenarios, which indicates that it fully understands the general knowledge of traffic behaviors and trajectories. 
Specifically, the multi-modal few-show CoT instructs the LMM to solve the above five stages of concrete scenario program generation step by step.
This process is given as follows:

\begin{enumerate}[leftmargin=*]

    \item \textbf{Instruct the LMM to learn about the goal and context of the task}. The CoT starts by informing the LMM of the task that generates concrete scenarios from input abstract scenario. This includes two prompts and their patterns are given in Table~\ref{first_prompt} (instruction and context).
    
    \item \textbf{Select the road for concrete scenario generation}. \name selects the road segment and retrieves the road information from the map file according to the road type specified in the abstract scenario. This contains the information for a group of lanes in the road, including lane ID, lane direction and lane length.
    
    \item \textbf{Guide the LMM to assign ego vehicle's initial position and destination on the road map}. Providing the road map (picture file), \name directs the LMM to determine two positions on the given road as the initial position and destination of the ego vehicle. By generating the example for the basic driving task \textit{follow lane} on the road (two positions on a lane along the lane direction), \name instructs the LMM that the ego vehicle's initial position and destination should perform the driving task specified in abstract scenario. The prompt pattern of this step is given in Table~\ref{first_prompt} (ego determination).
    
    \item \textbf{Instruct the LMM to divide the road}. \name instructs the LMM to divide the lanes of the road into a group of divisions relative to ego vehicle's initial position. The prompt pattern of this step is given in Table~\ref{first_prompt} (road divisions).
    
    \item \textbf{Guide the LMM to generate trajectories for NPC vehicles and pedestrians}. Based on the road divisions, \name guides the LMM to generate trajectories of participants' behaviors specified in the abstract scenario. It adds the example trajectory on the road for a basic participant behavior \textit{follow lane from left/right front} (a trajectory in front along the lane direction on the left/right lane), to make the LMM learn how to generate the trajectory according to the attributes of the participant behavior described in the abstract scenario. The prompt pattern of this step is given in Table~\ref{first_prompt} (participant trajectory).
    
    \item \textbf{Add the test scenario assertion statement}. \name provides the assertion template for the LMM, and promotes it to complete the assertion statements into the concrete scenario program. The assertion uses Signal Temporal Logic (STL) to monitor the ego vehicle: (i) whether it collides with other objects; (ii) whether the ego vehicle always keeps a safe distance from NPC vehicles and pedestrians; and (iii) whether the ego vehicle can arrive at its destination in time. An example is shown in lines 26-34 of List~\ref{program}. 
\end{enumerate}

\begin{table}[h]\footnotesize
\centering
\renewcommand\arraystretch{1.3}
\vspace{-5px}
\caption{Prompt patterns for concrete scenario generation.}
\vspace{-10px}
\begin{tabular}{c|l}
\hline
\textbf{Type}                                                    & \multicolumn{1}{c}{\textbf{Sample of prompt patterns}}                                                                                                                                                                                                                                \\ \hline
\begin{tabular}[c]{@{}c@{}}instruc-\\ tion\end{tabular}          & \begin{tabular}[c]{@{}l@{}}You are an expert in autonomous driving system (ADS) testing. We want you to generate \textless{}count\textgreater\\ test scenarios according to the input to challenge the ego vehicle (which connects to ADS).\end{tabular}                              \\ \hline
context                                                          & \begin{tabular}[c]{@{}l@{}}A scenario to test ADS is shown as follows: \textless{}A scenario program example\textgreater{}.\\ \textless{}Introduction of parameters for ego vehicle\textgreater{}. \textless{}Introduction of parameters for participants\textgreater{}.\end{tabular} \\ \hline
\begin{tabular}[c]{@{}c@{}}ego deter-\\ mination\end{tabular}    & \begin{tabular}[c]{@{}l@{}}On the input road, S and D are the examples of ego vehicle's initial position and destination\\ for "change lane" task. S is defined by ("lane\_222"$\rightarrow$10), D is defined by ("lane\_223"$\rightarrow$110).\end{tabular}                                                  \\ \hline
\begin{tabular}[c]{@{}c@{}}road\\ divisions\end{tabular}         & \begin{tabular}[c]{@{}l@{}}The road for test scenario generation is divided into \textless{}number\textgreater divisions: \textless{}{}\textgreater{}correspond to relative \\ positions according to the position of participant relative to the ego vehicle's initial position.\end{tabular}       \\ \hline
\begin{tabular}[c]{@{}c@{}}participant\\ trajectory\end{tabular} & \begin{tabular}[c]{@{}l@{}}On the input road, for an NPC vehicle that performs "follow lane" behavior (NPC's relative position\\ is "right front"), its waypoints are defined as:(("lane\_223"$\rightarrow$30, ,5),("lane\_223"$\rightarrow$100, ,8))\end{tabular}                                            \\ \hline
\end{tabular}
\vspace{-6px}
\label{first_prompt}
\end{table}

To test whether the multi-modal few-shot CoT can guide the LMM to generate the corresponding trajectories for all behaviors on different types of roads, we select the abstract scenarios extracted from the above 50 scenarios that contain various types of behaviors and roads, and test the LMM's performance on concrete scenario generation. The results show that by our multi-modal CoT prompting, \name can correctly generate trajectories for behaviors on different roads. Further evaluation is shown in Section~\ref{sec:rq1}.

\subsubsection{Scenario Inspection.} 
\label{sec:scenario-inspect}
As the generated concrete scenarios should adhere to realistic traffic, \name checks the feasibility of generated participants’ trajectories. This entails the guarantees that NPC vehicles operate in the correct position, directions and speeds, and do not violate traffic regulations. To achieve this, \name checks the feasibility of trajectories in the scenarios by the general constraints. For the scenario with infeasible trajectories, \name feeds it back to the LMM to generate a feasible scenario. The constraints in our consideration is listed as below: 

\begin{itemize}[leftmargin=*]
    \item \textbf{Constraints for each vehicle's heading and driving direction}: the heading and driving direction should be the same as the lane direction, and cannot move backward on the same lane.
    \item \textbf{Spatial constraint}: For each NPC vehicle, the initial position cannot be partially overlapped, and they are constrained to be at least 5 meters away from each other. 
    \item \textbf{Constraint for speed}: the speeds of vehicles should not exceed the speed limit of the road.
    \item \textbf{Temporal constraint}: the movement of each NPC vehicle's speed and position offset is structured sequentially, following the sequence of states of trajectories.
\end{itemize}

\subsection{Dual-Layer Search for Safety-Critical Semantic Equivalent Test Scenarios} 
\label{sec:method-search}

Based on the concrete scenarios generated from real traffic videos, \name searches for safety-critical scenarios and test the performance of the ADS under these traffic situation. To achieve this, we design a dual-layer optimization search technique, as described below. 

\begin{itemize}[leftmargin=*]
\item \textbf{Outer-Layer Optimization:} The outer-layer optimization refines scenario space to maximize the differences of ego vehicle's behaviors and human-driving behaviors for the same driving task in semantic equivalent test scenarios. 
\item \textbf{Inner-Layer Optimization:} Based on the discovered safety-violation scenario of the ADS, the inner-layer optimization explores variations in participants' trajectories to verify whether the ADS's safety violation is universal in the traffic situation of the abstract scenario.
\end{itemize}

During the optimization search, to ensure the semantic consistency of the generated test scenarios with the abstract scenario of the traffic scenario video, for the newly generated trajectories, \name checks their behaviors and feasibility by action specifications. Each generated trajectory consists of a sequence of waypoints, and each waypoint $w_i = (pos^i, vel^i)$ contains the position ($pos^i = (x^i,y^i)$) and velocity ($vel^i = (v^i_x, v^i_y)$). Based on the trajectory, \name first uses linear interpolation on the trajectory to identify its significant changes of motion states during driving, which is similar to the motion change identification in Section \ref{sec:motion-change-extraction}. For each motion segment of the ego vehicle's trajectory, \name recognizes its action by a set of specifications. Due to the page limit, we take the specifications of \textit{change left} as the example, and the other specifications can be found in our project website: \url{https://anonymous.4open.science/r/CRISER}.
\begin{enumerate}[leftmargin=*]

\item \textbf{Lane specification for \textit{change left} identification}: $pos^0 \in l_s, pos^F \in l_f, (s \not= f) \wedge (l_s,l_f \in R), (df(l_s) = df(l_f)) \wedge (-1<\sin(fd(pos^0, pos^F),df(l_s))<0)$,
where $pos^0$ is the position of the initial waypoint on the trajectory and $pos^F$ is the last waypoint. $l_s$ and $l_f$ are two lanes on the road $R$. $df$ represents the direction and $fd(A, B)$ is the angle from direction A to direction B. This specification specifies that the starting position and ending position are on two different lanes on the same road, and the directions of two lanes are the same. The ending position is on the left lane of starting position.

\item \textbf{Driving position specification for \textit{change left} identification}: $\forall i \in (0, F), pos^i \in l_s \bigcup l_f,$ $ pos^i.x \in [\min(pos^0.x, pos^F.x), \max(pos^0.x, pos^F.x)], pos^i.y \in [\min(pos^0.y, pos^F.y), \max(pos^0.y, $ $pos^F.y)]$, 
where $i$ represents the index of the waypoint in the trajectory segment. This specification specifies that the positions of intermediate waypoints of the trajectory segment are between the starting position and ending position.

\item \textbf{Driving direction specification for \textit{change left} identification}: $\forall i \in (0,F), arctan 2 (vel^i_y, vel^i_x) $ $ \in (90, 180)$. This specification specifies that the driving directions are to the left.

\item \textbf{Speed constraints:} $\forall i \in (0, F), (\lvert vel^i \lvert \le speed_{max}) \wedge (\lvert vel^{i+1} \lvert) - \lvert vel^i \lvert)/\Delta t < threshold_{c}$, where $speed_{max}$ is the speed limits of the road, and $threshold_c$ is the threshold for speed change during the actions except accelerating, decelerating and braking.

\end{enumerate}

\subsubsection{Outer-Layer Optimization.}
\label{sec:out-layer}

The layer is responsible for the global objective. \name evolves the concrete scenarios by maximizing the difference between human-driving behaviors in real-traffic scenarios and ego vehicles' behaviors in concrete scenarios semantically equivalent with the abstract scenario. It has the following steps.

The first step is to assess the behavior differences between the ego vehicle and human-driving in semantic equivalent scenarios. \name represents each of these two types of behaviors as an action sequence. The action sequence of human-driving $A_{h} = \{ b_1, b_2, ... b_n \}$ can be extracted from the abstract scenario. $b_n$ is a behavioral action (e.g., \textit{change left}, \textit{turn right}). The action sequence of the ego vehicle $A_{E} = \{ a_1, a_2, ... a_m \}$ is extracted based on the execution data of the semantic equivalent scenario. \name assesses the difference of these two sequences using the following deviation metric: 
\begin{equation}
    D (A_E, A_h) = \sum_{i=1, j=1}^{m, n} D(i, j), 
\end{equation}
where $D(i, j)$ is calculated by the \textit{Levenshtein Distance}.
In the concrete scenarios and corresponding traffic-video scenarios, the number of actions in human-driving sequence and ego vehicle sequence may be different and overlapped. The Levenshtein Distance calculates the difference between the two sequences by transforming one sequence into another through insertion, deletion, and replacement operations. For $A_{h} = \{ b_1, b_2, ... b_n \}$ and $A_{E} = \{ a_1, a_2, ... a_m \}$, the distance calculation adopts a recursive method as follows:
\begin{equation}
    D (i, j) = \left\{ 
                    \begin{array}{lccl}
                    0, & & if & i=0, j=0 \\
                    D(i-1, j-1), & & if & a_i = b_j  \\
                    min \left\{ \begin{array}{l}
                    D(i-1, j) + cost\_del(a_i), \\
                    D(i, j-1) + cost\_ins(b_j), \\
                    D(i-1, j-1) + cost\_rep(a_i, b_j),
                    \end{array}
                    \right. & & if & a_i \not= b_j
                    \end{array} 
                \right.
\end{equation}
where $cost\_del(a_m)$ represents the cost of the deletion operation that deletes $a_m$ from the action sequence $A_E$; $cost\_ins(b_n)$ represents the cost of the insertion operation that inserts $a_n$ into the action sequence $A_E$; $cost\_rep(a_m, b_n)$ represents the cost of the replacement operation that replaces $a_m$ with $b_n$ in the action sequence $A_E$. For example, $A_h=\{$\textit{follow lane, decelerate, change right, accelerate, cross}$\}$ and $A_h=\{$\textit{follow lane, brake, change right, accelerate, decelerate, cross}$\}$. $D(AD, HM)=$\textit{cost\_rep(brake, decelerate)+cost\_ins(decelerate)}.
The cost of insertion and deletion is defined as a constant $\lambda$, and the cost of replacement is determined by the types of $a_m$ and $b_n$. For instance, the cost of replacing $a_m$ and $b_n$ is high if they are  \textit{change left} and \textit{change right}, and low if they are \textit{change left} and \textit{brake}. 

To extract the action sequence of ego vehicle in the test scenario, \name records its actual driving trajectory during the scenario execution. Based on the trajectory of ego vehicle, \name identifies its action sequence by motion state change and action specifications mentioned above.

The second step is scenario mutation.
\name employs the feedback-guided fuzzer to search for the test scenarios where the behaviors of the ego vehicle are significantly different from that of human-driving in semantic equivalent scenarios. The fuzzer is employed based on the metric $D(A_h, A_E)$ ($A_h$ represents the action sequence of human-driving and $A_E$ represents the action sequence of ego vehicle in the semantic equivalent scenario) and the Gaussian mutation \cite{song2021dimension} of the participants' types, positions and speeds. 

To reveal the potential deficiencies of the ADS, \name discovers its safety violations in semantic equivalent scenarios, including collisions, traffic disruption, and traffic rule violations. When the safety-violation scenario is found, \name inputs it into the inner-layer optimization. 


\subsubsection{Inner-Layer Optimization.}
The layer is to verify whether the safety violation of the ADS is universal or occasional in this traffic situation. To do this, for each discovered safety-violation scenario, \name expands the trajectory coverage of the involved participants while maintaining the original semantics of the abstract scenario. Specifically, for the safety-violation scenario discovered by outer-layer optimization, \name incrementally explores variations into the trajectory of each participant, which generates diverse trajectories for each participant's behaviors to test the ego vehicle's driving performance. If the safety violation of the ADS persists across variations of the test scenario, it is considered as a potential deficiency of the ADS. For the safety-violation scenario $sf$ and the set of its variations $\mathbb{Z}$ where the safety violation of the ego vehicle occurs, the range of variations $RV_{sf}$ is evaluated by the maximal \textit{Euclidean distance} between participants' trajectories across $sf$ and a scenario in $\mathbb{Z}$. Euclidean distance is widely used to compute the similarity of the participant trajectories \cite{li2020av} across two test scenarios.
\begin{equation}
RV_{sf}= arg \max_{s} \sum_{s \in \mathbb{Z}} ED_{sf, s}, \quad ED_{sf,s}=\frac{\sum_{n=1}^l \sum_{m=1}^c TD_{sf^n, s^m}}{l*c},
\end{equation}
where $ED$ represents the Euclidean distances of two test scenarios and $TD$ represents the Euclidean distances of two participant trajectories, calculated as follows: 
\begin{equation}
TD_{s_i^n, s_j^m}=\sum_{k=0}^{\mu} \sqrt{(x_{s_i^n.k}-x_{s_j^m.k})^2+(y_{s_i^n.k}-y_{s_j^m.k})^2}
\end{equation}
$sf^n$ represents the $n$-th participant trajectory in $sf$. The number of NPC vehicles/pedestrians in $sf$ and $s$ are $l$ and $c$ respectively. $(x_{sf^n.k}, y_{sf^n.k})$ represents the position of the $k$-th waypoint of the n-th participant trajectory in $sf$. $\mu$ represents the number of waypoints in the participant trajectory $sf_i$. If $RV_{sf} \ge \mathbf{M}$, \name records these safety-violation scenarios which can be replayed to reproduce the safety violation of the ADS.

 


\section{Evaluation}
\label{sec:result and comparative}

In the evaluation, we mainly target the safety testing of industry-grade L4 ADS. In particular, we select the open-source full-stack ADS, Baidu Apollo \cite{apolloauto}, for the following reasons. (1) \textit{Representativeness}. Apollo community ranks among the top-four leading industrial ADS developers \cite{rank}, while the other three ADSs are not publicly released. (2) \textit{Practicality}. Apollo can be readily installed on vehicles for driving on public roads \cite{launch}. It has be commercialized for many real-world self-driving services \cite{selfhighway, baidutaxi}). (3) \textit{Advancement}. Apollo is actively and rapidly updated to include the latest features and functionalities. Our proposed method can be applied to test other ADSs as well.

To demonstrate the ability of \name, we apply it to Apollo 7.0 \cite{apolloauto}, which is widely used to control AVs on public roads. To evaluate the efficiency and effectiveness of \name, we answer the following research questions:

\begin{itemize}[leftmargin=*]
    \item \textbf{RQ1:} How effective is \name's traffic video-based scenario understanding and generation? 
    
    \item \textbf{RQ2:} How effective and efficient is \name in finding safety violations of Apollo in semantic equivalent scenarios?

\end{itemize}


\subsection{Experiment Settings}
We conducted the experiments on Ubuntu 20.04 with 500 GB memory, an Intel Core i7 CPU, and an NVIDIA GTX2080 TI. SORA-SVL \cite{sora-svl} (an end-to-end AV simulation platform that supports connection with Apollo) and San Francisco map are selected to execute the generated scenarios. During the experiments, all modules of Apollo are turned on, including perception, localization, prediction, routing, planning, and control modules.

We select the automotive video dataset, HRI Driving Dataset (HDD) \cite{ramanishka2018CVPR}, to generate traffic realistic scenarios.
HDD is a dataset to reflect driver behavior and various driving situations in real traffic, which includes 104 hours of safe human driving in the San Francisco collected using an on-vehicle recorder. The dataset encompasses various types of roads (straight road, intersection, T-junction) and participant behaviors (follow lane, change left/right, turn left/right, cross, accelerate, decelerate, brake, walk along, walk across, stand). We adopt GPT-4~\cite{gpt4v1} as the LMM because it is the current state-of-the-art LMM, widely known and easily accessible. 

For the test scenario description, we review the existing ADS testing Domain-Specific Languages (DSLs) that can describe scenarios to test Apollo \cite{jullien2009openscenario, fremont2019scenic, zhou2023specification}. We select AVUnit \cite{avunit} because it can accurately define and deterministically execute the motion trajectories of participants in test scenarios. In an AVUnit scenario program, each trajectory is defined as a sequence of states, and each state consists of the position, heading and speed. 


Some parameters of \name need to be defined: we set 
the interpolation parameter $\alpha$ as 0.2, as the key action extraction of \name is required to identify and retain the instantaneous changes of motion changes. The smaller value of $\alpha$ is sensitive to sudden changes and can more promptly reflect changes in actions. 
The $threshold_c$ is set as 0.5$m/s^2$ due to that: according to research on driving behavior \cite{salvucci2004two, neumann2011two}, changes in speed below 0.5$m/s^2$ during human driving are often considered steady and not recognized as obvious acceleration or deceleration action. We set $\mathbf{M}$ as 10 referencing to \cite{tian2022mosat, li2020av}, which indicates that different test scenarios with Euclidean distance greater than 20 can be used as classification criteria for different categories of scenarios.

\subsection{Experiment Design}
\noindent\textbf{For RQ1}, we use \name to understand the scenario semantics of traffic videos, and then generate executable test scenarios. To evaluate the effectiveness of scenario semantics understanding for traffic videos, we employ M-CPS \cite{zhang2023building} and $LEADE_E$ as the baselines. M-CPS is a model to extract effective information from accident videos. $CRISER$ directly splits the traffic video at 1s time-step and leverages GPT-4V to understand the semantics of the sequences of frames. We randomly select 100 different videos that encompass various types of roads, ego vehicle's driving tasks, participant types and behaviors. Then we run \name, M-CPS and $LEADE_E$ to understand scenario semantics of the same videos respectively.

To validate the effectiveness of \name in concrete scenario generation, we use CRISCO \cite{tian2022generating} and $LEADE_D$ as the baselines. CRISCO defines hundreds of constraints to generate concrete scenarios based on abstract scenarios by solving these constraints. $LEADE_D$ directly leverages GPT-4V to generate concrete scenarios from abstract scenarios on the given roads.
Next, we run \name, CRISCO and $LEADE_D$ to generate concrete scenarios based on the same abstract scenarios. 

Four of the authors independently analyze and cross-check the extracted information of each scenario video and the generated concrete scenarios of each abstract scenario. Another author is involved in a group discussion to resolve conflicts and reach agreements.

\noindent\textbf{For RQ2}, we randomly select 100 traffic scenarios from traffic video dataset, and uses \name to generate test scenarios to test Apollo. For the recorded safety-violation scenarios, we analyze the potential deficiencies and correct operations of modules in Apollo.
To evaluate the effectiveness and efficiency of \name in discovering safety violations of Apollo in semantic equivalent scenarios, we build a variant of \name and employ a state-of-the-art safety-critical scenario generation technique as baselines for comparison: $LEADE_r$ and M-CPS \cite{zhang2023building}, which can generate safety-violation scenarios for Apollo based on traffic videos. M-CPS builds test scenarios by reproducing and mutating scenarios from traffic videos. $LEADE_r$ is built by randomly searching participants' parameters within the space of preserving the scenario semantics of traffic videos. 
We run \name, M-CPS and $LEADE_r$ based on the same traffic videos to generate the same amount of test scenarios, and compare their effectiveness and efficiency from the following aspects:
\begin{itemize}[leftmargin=*]
    \item How many types of Apollo safety violations are found by \name and baselines in their generated test scenarios?
    \item How many different traffic videos are leveraged by \name and baselines to expose safety violations of Apollo? 
\end{itemize}
Note that to account for the randomness of \name's dual-layer optimization search, this experiment is repeated 5 times.


\subsection{Result Analysis: RQ1}
\label{sec:rq1}
For the accuracy of scenario understanding of traffic videos, we evaluate it by calculating the information extraction accuracy of elements in scenarios. The elements are divided into four categories: road, ego driving task, participants, relative positions. According to the abstract scenario, each category of elements include several attributes. The attributes of road include road type and traffic signal. The attributes of ego driving task include the behaviors of ego vehicle. The attributes of participant include participant types and participant behaviors. The attributes of relative positions include the initial positions and destinations of participants relative to ego vehicle. Here the understanding of an element in a scenario is accurate if all attributes of the element are extracted correctly. The information extraction accuracy of the elements of category $c$ is represented as $\mathbf{SUA}_c$, calculated as follows:
{\setlength\abovedisplayskip{0cm}
\setlength\belowdisplayskip{0.1cm}
\begin{equation}
    \mathbf{SUA}_c = \frac{1}{\Vert \mathbb{TV} \Vert \times \Vert A_i^c\Vert} \sum_{i=1}^{\Vert \mathbb{TV}\Vert} \sum_{j=1}^{\Vert A_i^c\Vert} \mathbbm{1} \left( \bigwedge_{\forall ar \in A^c_i.j} M(ar, \mathbb{TV}.i) \right) 
\end{equation}}
where $\mathbb{TV}$ represents the set of traffic scenarios and $A_i^c$ represents all extracted elements of category $c$ from $i$-th scenario. $\mathbbm{1}$ is an indicator function mapping a boolean condition to a value in \{0, 1\}: if the condition is true, it returns 1; otherwise, it returns 0. $\bigwedge$ means logical \textit{AND}. $A^c_i.j$ represents the extracted attribute values of the $j$-th element of category $c$ in the $i$-th scenario. $M$ is a function that evaluates whether the extracted value of the attribute conforms to the ground truth of corresponding traffic scenario. 

\begin{table}[h]
\centering
\small
\vspace{-5px}
\caption{The accuracy of scenario understanding of traffic videos}
\vspace{-5px}
\renewcommand\arraystretch{1.3}
\begin{tabular}{c|c|c|c|c}
\hline
\textbf{Element Category}      & \textbf{Road} & \textbf{\begin{tabular}[c]{@{}c@{}}Ego Task\end{tabular}} & \textbf{\begin{tabular}[c]{@{}c@{}}Participant\end{tabular}} & \textbf{\begin{tabular}[c]{@{}c@{}}Relative Positions\end{tabular}} \\ \hline
\name     & 97\%          & 96\%                                                        & 92.1\%                                                                  & 86.1\%                                                               \\ \hline
$LEADE_E$ & 76\%          & 73\%                                                        & 69.0\%                                                                  & 54.2\%                                                               \\ \hline
M-CPS     & 89\%          & 84\%                                                        & 61.6\%                                                                  & 58.3\%                                                               \\ \hline
\end{tabular}
\vspace{-5px}
\label{extract_result}
\end{table}

Table~\ref{extract_result} shows the accuracy of \name, $LEADE_E$ and M-CPS in extracting information from traffic videos. On all extracted attributes, the accuracy of \name is high and performs better than the two baselines. The relatively low accuracy of $LEADE_E$ demonstrates the effectiveness of our information extraction prompt. M-CPS performs well in extracting static information (e.g., road structure) from traffic scenarios. However, for the complex dynamic information (e.g., participant behaviors), the extraction accuracy of M-CPS significantly decreases. The same problem also exists for CRISER, but it can almost accurately extract key information of traffic scenarios. 

For the correctness of concrete scenario generation, we use \textit{Concrete Scenario Correctness} $\mathbf{CSC}$ as the metric. Here a generated concrete scenario is correct if all elements during the execution of the scenario conform to the scenario semantics of corresponding traffic video. The calculation of $\mathbf{CSC}$ is given as follows:
{\setlength\abovedisplayskip{0cm}
\setlength\belowdisplayskip{0.1cm}
\begin{equation}
    \mathbf{CSC} = \frac{1}{\Vert \mathbb{S} \Vert} \sum_{i=1}^{\Vert \mathbb{S}\Vert} \mathbbm{1} \left( \bigwedge_{\forall sr \in \mathbb{S}.i } Sim(sr, \mathbb{AS}.i) \right) 
\end{equation}}
where $\mathbb{S}$ represents the set of generated concrete scenarios and $\mathbb{AS}$ represents the set of abstract scenarios. $p$ represents the execution result of an element in the concrete scenario. $Sim$ is to determine that during the execution of concrete scenario, whether the element in it conforms to the semantics of corresponding abstract scenario. 

\begin{table}[h]
\centering
\small
\vspace{-5px}
\caption{The correctness of concrete scenarios generation}
\vspace{-5px}
\renewcommand\arraystretch{1.3}
\begin{tabular}{c|c|c|c}
\hline
\textbf{Road Type}      & \textbf{Straight} & \textbf{Intersection} & \textbf{T-junction} \\ \hline
\textbf{CRISER}     & 95.5\%            & 91.4\%                & 90\%                \\ \hline
\textbf{$LEADE_D$} & 57.8\%            & 31.4\%                & 30\%                \\ \hline
\textbf{CRISCO}    & 93.3\%            & 85.7\%                & 75\%                \\ \hline
\end{tabular}
\vspace{-5px}
\label{concrete_generation}
\end{table}

Due to the varying complexity of scenarios on different road types, there are differences in the correctness of concrete scenario generation. We conduct statistical analysis about the correctness of concrete scenario generation on each type of road. Table~\ref{concrete_generation} shows the results. The concrete scenario generation correctness of $LEADE_D$ is low, which indicates that it's infeasible to directly leverage the LMM to generate concrete scenarios from abstract scenarios on the given roads. The correctness of \name is also higher than that of CRISCO, moreover, \name does not need to define the large set of constraints. The high correctness of \name demonstrates the effectiveness of the multi-modal few-shot CoT that guides the LMM to generate concrete scenarios according to abstract scenarios.

\subsection{Result Analysis: RQ2}
In each run, \name generates 10 test scenarios for each traffic video. 
To analyze the safety-violation scenarios caused by ego vehicle, for each recorded safety-violation scenario, we check whether the safety violation of Apollo is caused by the illegal actions of participants in the scenario. If true, we exclude it.
In the 1000 generated test scenarios, on average, 167 (min 155 and max 176) out of them expose safety violations of Apollo. 

To better analyze and clarify the potential deficiencies of Apollo, for each safety-violation scenario, we identify the \textit{essential participants} that cause ego vehicle's safety violation. To identify essential participants of each safety-violation scenario, \name first replays it by removing its participants one by one, and checking whether the safety violation of ego vehicle still occurs. If true, the participant is not essential for the safety violation. We analyze and classify the safety-violation scenarios based on the semantics of essential participants. The results are shown as Table~\tabref{safety-violation-type}, where EV represents the ego vehicle. NV represents NPC vehicles (2 NVs represent two NPC vehicles), and P represents pedestrians.
As the page limitation, we select one to explain in the following (EV represents the ego vehicle). The illustration for other types of safety violations can be found in the safety violation folder of \url{https://anonymous.4open.science/r/CRISER}.



\begin{table}[t]
\footnotesize
\vspace{-5px}
\caption{The discovered safety violations of Apollo by \name.}
\vspace{-5px}
\resizebox{\linewidth}{!}{
\begin{tabular}{|c|c|c|ccc|c|}
\hline
\multirow{2}{*}{\textbf{No}} & \multirow{2}{*}{\textbf{\begin{tabular}[c]{@{}c@{}}Road\\ type\end{tabular}}} & \multirow{2}{*}{\textbf{\begin{tabular}[c]{@{}c@{}}Driving\\ task\end{tabular}}} & \multicolumn{3}{c|}{\textbf{Participant}}                                                                                                                                                                                                                 & \multirow{2}{*}{\textbf{Driving Error of Apollo}}                                                                                                                                                  \\ \cline{4-6}
                             &                                                                               &                                                                                  & \multicolumn{1}{c|}{\textbf{type}}                                      & \multicolumn{1}{c|}{\textbf{relative positions}}                                                    & \textbf{behaviors}                                                        &                                                                                                                                                                                                    \\ \hline
1                            & \begin{tabular}[c]{@{}c@{}}Inters-\\ ection\end{tabular}                      & \begin{tabular}[c]{@{}c@{}}Turn\\ right\end{tabular}                             & \multicolumn{1}{c|}{2 NV}                                               & \multicolumn{1}{c|}{left vertical}                                                                  & \begin{tabular}[c]{@{}c@{}}cross\\ cross+acc-\\ lerate\end{tabular}       & \begin{tabular}[c]{@{}c@{}}Misidentifying speed of NPC vehicles\\ driving one after one as the same,\\ leanding to misjudgement of the\\ acceleration of later vehicles\end{tabular}               \\ \hline
2                            & \begin{tabular}[c]{@{}c@{}}Straight\\ road\end{tabular}                       & \begin{tabular}[c]{@{}c@{}}Change\\ left/right\end{tabular}                      & \multicolumn{1}{c|}{2 NV}                                               & \multicolumn{1}{c|}{\begin{tabular}[c]{@{}c@{}}left/right front\\ left/right behind\end{tabular}}   & \begin{tabular}[c]{@{}c@{}}follow lane\\ accelerate\end{tabular}          & \begin{tabular}[c]{@{}c@{}}Ignore acceleration of NPC vehicle\\ behind on adjacent lane,leading to\\ not keeping the safe distance\end{tabular}                                                    \\ \hline
3                            & \begin{tabular}[c]{@{}c@{}}Interes-\\ ection\end{tabular}                     & Cross                                                                            & \multicolumn{1}{c|}{2 NV}                                               & \multicolumn{1}{c|}{left vertical}                                                                  & \begin{tabular}[c]{@{}c@{}}cross\\ follow lane\\ \&turn left\end{tabular} & \begin{tabular}[c]{@{}c@{}}Misidentifying vehicles driving side\\ by side on vertical lanes as one,ignor-\\ ing behavior changes of later vehicle\end{tabular}                                     \\ \hline
4                            & \begin{tabular}[c]{@{}c@{}}Interes-\\ ection\end{tabular}                     & \begin{tabular}[c]{@{}c@{}}Turn\\ left\end{tabular}                              & \multicolumn{1}{c|}{1 NV}                                               & \multicolumn{1}{c|}{left vertical}                                                                  & turn left                                                                 & \begin{tabular}[c]{@{}c@{}}When turning at connection area of\\ intersection, there is delay of Apollo\\ in processing and responding to\\ the right of way of NPC vehicles\end{tabular}           \\ \hline
5                            & \begin{tabular}[c]{@{}c@{}}Straight\\ road\end{tabular}                       & \begin{tabular}[c]{@{}c@{}}Drive\\ through\end{tabular}                          & \multicolumn{1}{c|}{2 NV}                                               & \multicolumn{1}{c|}{\begin{tabular}[c]{@{}c@{}}ahead\\ left/right front\end{tabular}}               & \begin{tabular}[c]{@{}c@{}}stop \&\\ decelerate\end{tabular}              & \begin{tabular}[c]{@{}c@{}}Disable to change lane again\\ during driving\end{tabular}                                                                                                              \\ \hline
6                            & \begin{tabular}[c]{@{}c@{}}T-\\ junction\end{tabular}                         & \begin{tabular}[c]{@{}c@{}}Turn\\ right\end{tabular}                             & \multicolumn{1}{c|}{1 NV}                                               & \multicolumn{1}{c|}{left vertical}                                                                  & \begin{tabular}[c]{@{}c@{}}cross \&\\ stop\end{tabular}                   & \begin{tabular}[c]{@{}c@{}}Disable to adjust route to another\\ incoming lane during turning\end{tabular}                                                                                          \\ \hline
\multirow{2}{*}{7}           & \multirow{2}{*}{\begin{tabular}[c]{@{}c@{}}Inters-\\ ection\end{tabular}}     & \begin{tabular}[c]{@{}c@{}}Turn\\ left\end{tabular}                              & \multicolumn{1}{c|}{2 NV}                                               & \multicolumn{1}{c|}{opposite}                                                                       & cross                                                                     & \multirow{2}{*}{\renewcommand{\arraystretch}{0.8}\begin{tabular}[c]{@{}c@{}}Inaccurate calculation of distance and\\ motion status of two consecutive NPC\\ vehicles passing through an intersection\end{tabular}}                  \\ \cline{3-6}
                             &                                                                               & Cross                                                                            & \multicolumn{1}{c|}{2 NV}                                               & \multicolumn{1}{c|}{left vertical}                                                                  & cross                                                                     &                                                                                                                                                                                                    \\ \hline
8                            & \begin{tabular}[c]{@{}c@{}}Inters-\\ ection/T-\\ juntion\end{tabular}         & Cross                                                                            & \multicolumn{1}{c|}{\begin{tabular}[c]{@{}c@{}}1 NV\\ 1 P\end{tabular}} & \multicolumn{1}{c|}{\begin{tabular}[c]{@{}c@{}}left/right front\\ left/right vertical\end{tabular}} & \begin{tabular}[c]{@{}c@{}}follow lane\\ cross\end{tabular}               & \begin{tabular}[c]{@{}c@{}}When participants in front perform other\\ abnormally violent actions (e.g., \\ emergency braking), EV is lack of\\ prediction of potential dangers nearby\end{tabular} \\ \hline
9                            & \begin{tabular}[c]{@{}c@{}}Straight\\ road\end{tabular}                       & \begin{tabular}[c]{@{}c@{}}Drive\\ through\end{tabular}                          & \multicolumn{1}{c|}{2 NV}                                               & \multicolumn{1}{c|}{ahead}                                                                          & decelerate                                                                & \begin{tabular}[c]{@{}c@{}}Misidentifying the two slow-speed\\ NPC vehicles ahead as static\\ objects with abnormal movement\end{tabular}                                                          \\ \hline
\multirow{2}{*}{10}          & \multirow{2}{*}{\begin{tabular}[c]{@{}c@{}}Inters-\\ ection\end{tabular}}     & \begin{tabular}[c]{@{}c@{}}Turn\\ right\end{tabular}                             & \multicolumn{1}{c|}{truck}                                              & \multicolumn{1}{c|}{left vertical}                                                                  & cross                                                                     & \multirow{2}{*}{\renewcommand{\arraystretch}{0.8}\begin{tabular}[c]{@{}c@{}}Lack of response to large vehicle\\ dimensions when EV accelerates,leading\\ to no adjustment for lateral spacing\end{tabular}}                         \\ \cline{3-6}
                             &                                                                               & Cross                                                                            & \multicolumn{1}{c|}{truck}                                              & \multicolumn{1}{c|}{right vertical}                                                                 & turn right                                                                &                                                                                                                                                                                                    \\ \hline
\end{tabular}}
\label{safety-violation-type}
\vspace{-5px}
\end{table}





\begin{figure}[h]
  \centering
  \includegraphics[scale=0.4]{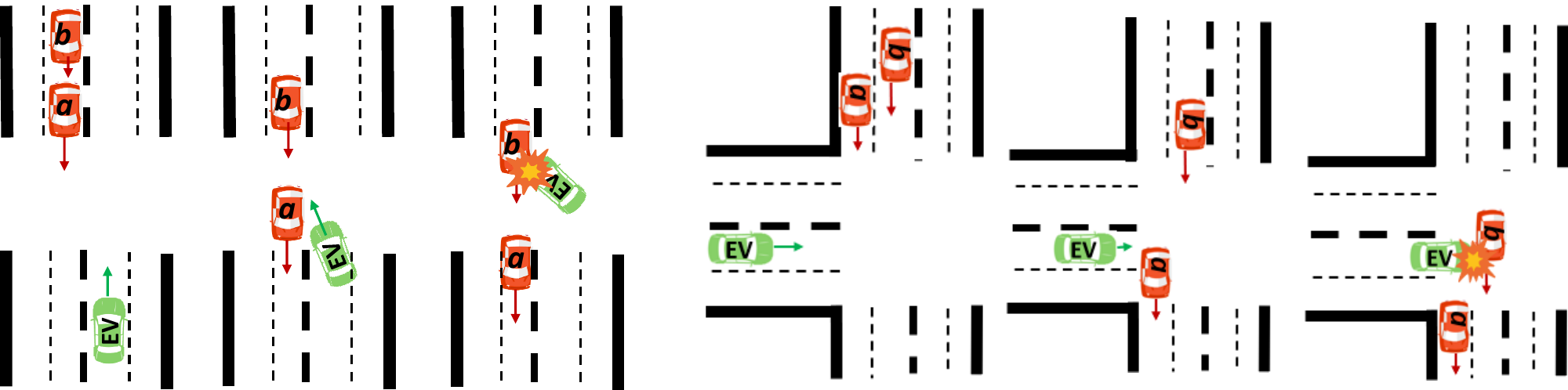}
  \vspace{-5px}
  \caption{Two examples of safety violation 7}
  \label{running3}
  \vspace{-10px}
  \end{figure} 

\textbf{Examples of safety violation 7:} As shown in \figref{running3}, in the left scenario, the driving task of human driver and EV is \textit{turn left}. Vehicle $a$ is crossing the intersection from the opposite side lanes, and vehicle $b$ is following $a$ to cross the intersection. In the real traffic video, the human driver decelerates to wait for vehicle $b$ to pass by, and then accelerates to finish the \textit{turn left}. In the test scenario of Apollo, before EV starts to turn at the entrance of junction, Apollo identifies the right of way of $a$ and stops to let $a$ pass. When vehicle $a$ passes the junction, EV continues to turn left due to wrongly calculation of the distance and speed of $b$, causing collision to $b$. The safety violation of the right scenario is caused by the same error of EV. The driving task of human driver and EV is \textit{cross}. The human driver either crosses the connection area with acceleration to turn left before vehicle $a$ approaching, or waits for vehicle $b$ to pass by before turning left. EV waits for vehicle $a$ to pass by but ignoring vehicle $b$ approaching, leading to the collision.



\begin{table}[h]\small
\centering
\vspace{-5px}
\caption{Comparison results of \name and baselines}
\vspace{-5px}
\renewcommand\arraystretch{1.3}
\begin{tabular}{|c|c|ccc|ccc|}
\hline
\multirow{2}{*}{\textbf{Approach}} & \multirow{2}{*}{\textbf{\begin{tabular}[c]{@{}c@{}}Types\\ of SVs\end{tabular}}} & \multicolumn{3}{c|}{\textbf{Number of SVs}}                                    & \multicolumn{3}{c|}{\textbf{Number of TVs to find SVs}}                    \\ \cline{3-8} 
                                   &                                                                                  & {\textbf{min}} & \multicolumn{1}{c|}{\textbf{max}} & \textbf{avg} & \multicolumn{1}{c|}{\textbf{\quad min\quad \quad}} & \multicolumn{1}{c|}{\textbf{ \quad max\quad  \quad}} & \textbf{\quad avg\quad} \\ \hline
\name                               & 10                                                                               & \multicolumn{1}{c|}{105}          & \multicolumn{1}{c|}{176}          & 143          & \multicolumn{1}{c|}{66}           & \multicolumn{1}{c|}{81}           & 74           \\ \hline
$LEADE_R$                         & 5                                                                                & \multicolumn{1}{c|}{59}           & \multicolumn{1}{c|}{101}           & 79           & \multicolumn{1}{c|}{48}           & \multicolumn{1}{c|}{62}           & 54           \\ \hline
M-CPS                              & 3                                                                                & \multicolumn{1}{c|}{21}           & \multicolumn{1}{c|}{55}           & 38           & \multicolumn{1}{c|}{18}           & \multicolumn{1}{c|}{31}           & 25           \\ \hline
\end{tabular}
\label{comparison}
\vspace{-5px}
\end{table}

\begin{figure}[h]
    \centering
    \begin{subfigure}[b]{0.36\linewidth}
        \centering
        \includegraphics[width=\linewidth]{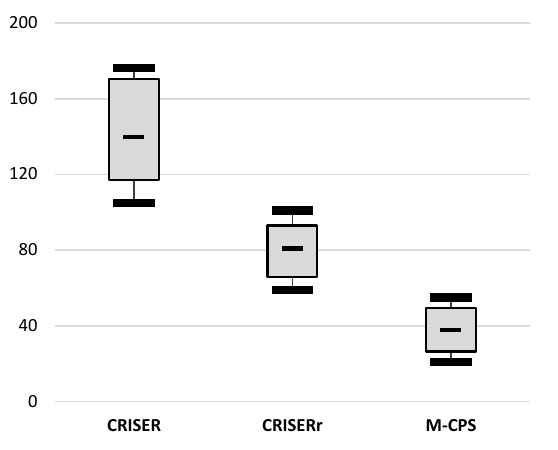}
        \caption{The number of discover SVs of Apollo}
        \label{sub1}
    \end{subfigure}
    \begin{subfigure}[b]{0.36\linewidth}
        \centering
        \includegraphics[width=\linewidth]{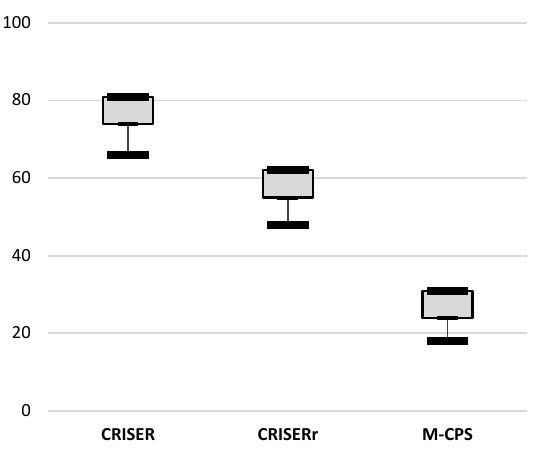}
        \caption{The number of found TVs to find SVs}
        \label{sub2}
    \end{subfigure}
    \vspace{-5px}
    \caption{The running results of \name, $LEADE_r$ and M-CPS}
    \label{comparison_box}
    \vspace{-10px}
\end{figure}

Table~\ref{comparison} and Figure~\ref{comparison_box} illustrate the comparison results of \name to baselines. SV is the abbreviation for ``safety violation'' and TV is the abbreviation for ``traffic video''. In each run, $LEADE_r$ generates 10 test scenarios for each traffic video by random sampling of parameters within the range of not changing the scenario semantics, and M-CPS generates 10 test scenarios for each traffic video by mutation algorithm for finding collision of the ego vehicle. \name can discover 10 distinct types of safety violations of Apollo. Based on the 100 traffic videos, \name utilizes 74 of them to discover safety violations of Apollo. $LEADE_r$ can discover 6 types of safety violations of Apollo, and M-CPS can find 3 types of safety violations of Apollo. On average, $LEADE_r$ generates 79 (min 59 and max 101) safety-violation scenarios, which are searched from 54 (min 48 and max 62) traffic videos. 
M-CPS generates 38 (min 21 and max 55) safety-violation scenarios, which are mutated from 25 (min 18 and max 31) traffic videos.
Compared with the two baselines, \name can effectively leverage more traffic videos to find more types of safety violations of Apollo. Furthermore, the performance of $LEADE_r$ is better than that of M-CPS, which indicates that \name's traffic video-based concrete scenario generation is helpful to discover ADS's safety violations in real traffic scenarios. 

\section{THREATS TO VALIDITY}
\noindent\textbf{Dataset Selection.} One primary threat to validity is the selection of the real-traffic video dataset. We select Honda Scenes \cite{narayanan2019dynamic}, a large-scale dataset that contains 80 hours of diverse high-quality driving video data clips. They are collected by vehicle cameras, encompassing various types of roads and participant behaviors.
With the growing popularity of vehicle cameras, the videos generated directly from vehicle cameras are considered one of the most reliable methods for traffic data collection \cite{miovision}.
We believe \name can be readily applied to other datasets of traffic video recordings.

\noindent\textbf{Count for Randomness.} Another threat to validity comes from the stochastic nature of the optimization search in \name. To count for its randomness, we conducted the experiment for RQ2 five times, and the results of each run exhibited slight variations. The running time of each experiment is long enough, and there is little difference among the experimental results of 5 runs. We provide the statistics and distributions of comparison aspects. Therefore, the experimental results can demonstrate the ability of \name in comparison to selected baselines. 

\section{Conclusion}
In this paper, we propose a novel approach, \name, that automatically generates abstract and concrete scenarios of real-traffic videos, and discovers safety violations of the ADS in scenarios with the semantics as real-traffic videos where human-driving works safely. \name leverages the LMM's capability in image understanding and program reasoning by motion change-based key frame extraction and multi-modal few-shot CoT, to generate abstract and concrete scenarios from traffic videos. Based on these scenarios, \name utilizes dual-layer search to maximize the differences between ego vehicle's behaviors and human-driving behaviors in semantically consistent scenarios, and verify the universality of the exposed safety violations of the ego vehicle. Experimental results show that \name can accurately generate abstract and concrete scenarios from traffic videos, and effectively discover more types of safety violations of the ADS.

\balance
\bibliographystyle{ACM-Reference-Format}

\begin{thebibliography}{57}


\ifx \showCODEN    \undefined \def \showCODEN     #1{\unskip}     \fi
\ifx \showDOI      \undefined \def \showDOI       #1{#1}\fi
\ifx \showISBNx    \undefined \def \showISBNx     #1{\unskip}     \fi
\ifx \showISBNxiii \undefined \def \showISBNxiii  #1{\unskip}     \fi
\ifx \showISSN     \undefined \def \showISSN      #1{\unskip}     \fi
\ifx \showLCCN     \undefined \def \showLCCN      #1{\unskip}     \fi
\ifx \shownote     \undefined \def \shownote      #1{#1}          \fi
\ifx \showarticletitle \undefined \def \showarticletitle #1{#1}   \fi
\ifx \showURL      \undefined \def \showURL       {\relax}        \fi
\providecommand\bibfield[2]{#2}
\providecommand\bibinfo[2]{#2}
\providecommand\natexlab[1]{#1}
\providecommand\showeprint[2][]{arXiv:#2}

\bibitem[nvi({[n.\,d.]})]%
        {nvidia_strive}
 \bibinfo{year}{[n.\,d.]}\natexlab{}.
\newblock \bibinfo{booktitle}{\emph{Automatically generating simulation accident scenarios for safe and scalable autonomous vehicle testing}}.
\newblock
\urldef\tempurl%
\url{http://nvidia.zhidx.com/content-6-3026.html}
\showURL{%
Retrieved July 23, 2024 from \tempurl}


\bibitem[sel({[n.\,d.]})]%
        {selfhighway}
 \bibinfo{year}{[n.\,d.]}\natexlab{}.
\newblock \bibinfo{booktitle}{\emph{Autoware Self-driving Vehicle on a Highway}}.
\newblock
\urldef\tempurl%
\url{https://www.youtube.com/watch?v=npQMzH3jd8}
\showURL{%
Retrieved Sepetem 1, 2023 from \tempurl}


\bibitem[avu({[n.\,d.]})]%
        {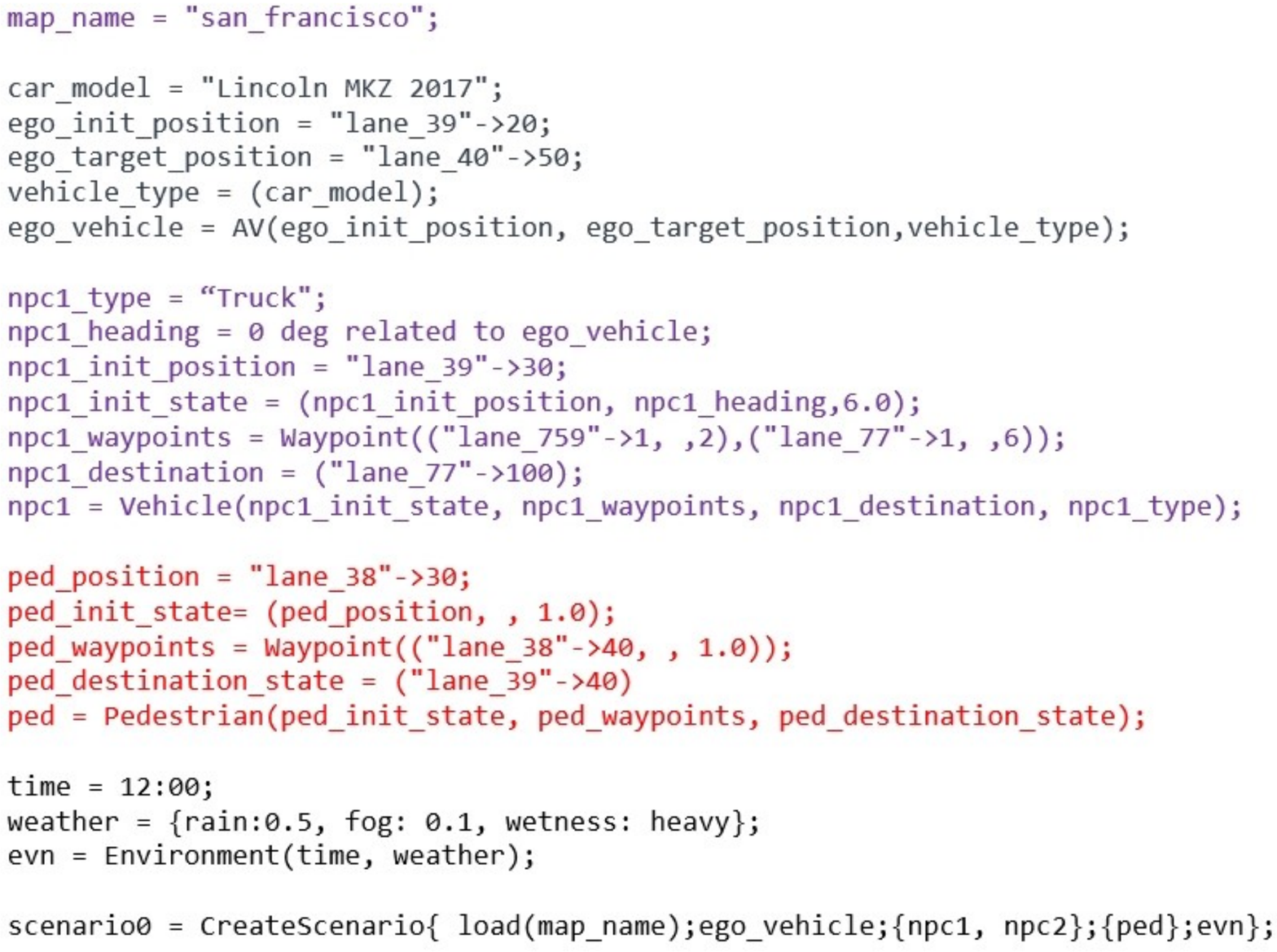}
 \bibinfo{year}{[n.\,d.]}\natexlab{}.
\newblock \bibinfo{booktitle}{\emph{AVUnit’s documentation}}.
\newblock
\urldef\tempurl%
\url{https://avunit.readthedocs.io/en/latest/Introduction_to_AVUnit.html}
\showURL{%
Retrieved April 12, 2024 from \tempurl}


\bibitem[bai({[n.\,d.]})]%
        {baidutaxi}
 \bibinfo{year}{[n.\,d.]}\natexlab{}.
\newblock \bibinfo{booktitle}{\emph{Baidu Launches Public Robotaxi Trial Operation}}.
\newblock
\urldef\tempurl%
\url{https://www.globenewswire.com/news-release/2019/09/26/1921380/0/ en/Baidu-Launches-Public-Robotaxi-Trial-Operation.html}
\showURL{%
Retrieved April 1, 2024 from \tempurl}


\bibitem[lau({[n.\,d.]})]%
        {launch}
 \bibinfo{year}{[n.\,d.]}\natexlab{}.
\newblock \bibinfo{booktitle}{\emph{Baidu launches their open platform for autonomous cars–and we got to test it}}.
\newblock
\urldef\tempurl%
\url{https://technode.com/2017/07/05/baidu-apollo-1-0-auto nomous-cars-we-test-it/}
\showURL{%
Retrieved April 1, 2024 from \tempurl}


\bibitem[ran({[n.\,d.]})]%
        {rank}
 \bibinfo{year}{[n.\,d.]}\natexlab{}.
\newblock \bibinfo{booktitle}{\emph{Navigant Research Names Waymo, Ford Autonomous Vehicles, Cruise, and Baidu the Leading Developers of Automated Driving Systems}}.
\newblock
\urldef\tempurl%
\url{https://www.businesswire.com/news/home/20200407005119/en/Navigant-Research-Names-Waymo-Ford-Autonomous-Vehicles}
\showURL{%
Retrieved April 1, 2024 from \tempurl}


\bibitem[mio({[n.\,d.]})]%
        {miovision}
 \bibinfo{year}{[n.\,d.]}\natexlab{}.
\newblock \bibinfo{booktitle}{\emph{Trusted Data for Mobility Planning-Portable Data Collection}}.
\newblock
\urldef\tempurl%
\url{https://miovision.com/solutions/data-collection-traffic-studies/}
\showURL{%
Retrieved July 28, 2024 from \tempurl}


\bibitem[way({[n.\,d.]})]%
        {waymo}
 \bibinfo{year}{[n.\,d.]}\natexlab{}.
\newblock \bibinfo{booktitle}{\emph{WAYMO's virtual world to test self-driving cars: Simulation City}}.
\newblock
\urldef\tempurl%
\url{https://www.d1ev.com/news/jishu/150890}
\showURL{%
Retrieved July 23, 2024 from \tempurl}


\bibitem[apo(2013)]%
        {apolloauto}
 \bibinfo{year}{2013}\natexlab{}.
\newblock \bibinfo{booktitle}{\emph{An open autonomous driving platform}}.
\newblock
\urldef\tempurl%
\url{https://github.com/ApolloAuto/apollo}
\showURL{%
Retrieved March 16, 2022 from \tempurl}


\bibitem[cha(2022)]%
        {chain_report}
 \bibinfo{year}{2022}\natexlab{}.
\newblock \bibinfo{booktitle}{\emph{Autonomous Driving Simulation Industry Chain Report (Foreign Companies)}}.
\newblock \bibinfo{publisher}{Research and Markets}.
\newblock


\bibitem[nht(2022)]%
        {nhtsa_files}
 \bibinfo{year}{2022}\natexlab{}.
\newblock \bibinfo{booktitle}{\emph{NHTSA}}.
\newblock
\urldef\tempurl%
\url{https://www.nhtsa.gov/sites/nhtsa.gov/files/811731.pdf}
\showURL{%
Retrieved May 11, 2022 from \tempurl}


\bibitem[sor(2023)]%
        {sora-svl}
 \bibinfo{year}{2023}\natexlab{}.
\newblock \bibinfo{booktitle}{\emph{SORA-SVL Simulator}}.
\newblock
\urldef\tempurl%
\url{https://github.com/YuqiHuai/SORA-SVL}
\showURL{%
Retrieved July 30, 2024 from \tempurl}


\bibitem[gpt(2024a)]%
        {gpt4-corner2}
 \bibinfo{year}{2024}\natexlab{a}.
\newblock \bibinfo{booktitle}{\emph{Five consecutive tests of GPT-4V's recognization in autonomous driving scenarios}}.
\newblock
\urldef\tempurl%
\url{https://mp.weixin.qq.com/s?__biz=MzIzNjc1NzUzMw==&mid=2247699188&idx=1&sn=e4b7957166950a52a4be69cd809cf1dd&scene=21#wechat_redirect}
\showURL{%
Retrieved July 28, 2024 from \tempurl}


\bibitem[gpt(2024b)]%
        {gpt4v1}
 \bibinfo{year}{2024}\natexlab{b}.
\newblock \bibinfo{booktitle}{\emph{GPT-4V(ision) System Card}}.
\newblock
\urldef\tempurl%
\url{https://cdn.openai.com/papers/GPTV_System_Card.pdf}
\showURL{%
Retrieved July 28, 2024 from \tempurl}


\bibitem[gpt(2024c)]%
        {gpt4v2}
 \bibinfo{year}{2024}\natexlab{c}.
\newblock \bibinfo{booktitle}{\emph{GPT-4V(ision) technical work and authors}}.
\newblock
\urldef\tempurl%
\url{https://cdn.openai.com/contributions/gpt-4v.pdf}
\showURL{%
Retrieved July 28, 2024 from \tempurl}


\bibitem[gpt(2024d)]%
        {gpt4-corner}
 \bibinfo{year}{2024}\natexlab{d}.
\newblock \bibinfo{booktitle}{\emph{GPT-4V's answer for autonomous driving corner case recognition}}.
\newblock
\urldef\tempurl%
\url{https://mp.weixin.qq.com/s/IV1BXmRCFwQs2CNXDknA8Q}
\showURL{%
Retrieved July 28, 2024 from \tempurl}


\bibitem[gpt(2024e)]%
        {gpt_quality}
 \bibinfo{year}{2024}\natexlab{e}.
\newblock \bibinfo{booktitle}{\emph{How does the prompt affect the quality of responses generated by ChatGPT?}}
\newblock
\urldef\tempurl%
\url{https://typeset.io/questions/how-does-the-prompt-affect-the-quality-of-responses-9zj2tuek2n}
\showURL{%
Retrieved July 30, 2024 from \tempurl}


\bibitem[Almaskati et~al\mbox{.}(2023)]%
        {almaskati2023autonomous}
\bibfield{author}{\bibinfo{person}{Deema Almaskati}, \bibinfo{person}{Sharareh Kermanshachi}, {and} \bibinfo{person}{Apurva Pamidimukkula}.} \bibinfo{year}{2023}\natexlab{}.
\newblock \showarticletitle{Autonomous vehicles and traffic accidents}.
\newblock \bibinfo{journal}{\emph{Transportation research procedia}}  \bibinfo{volume}{73} (\bibinfo{year}{2023}), \bibinfo{pages}{321--328}.
\newblock


\bibitem[Arrieta et~al\mbox{.}(2017)]%
        {arrieta2017search}
\bibfield{author}{\bibinfo{person}{Aitor Arrieta}, \bibinfo{person}{Shuai Wang}, \bibinfo{person}{Urtzi Markiegi}, \bibinfo{person}{Goiuria Sagardui}, {and} \bibinfo{person}{Leire Etxeberria}.} \bibinfo{year}{2017}\natexlab{}.
\newblock \showarticletitle{Search-based test case generation for cyber-physical systems}. In \bibinfo{booktitle}{\emph{2017 IEEE Congress on Evolutionary Computation (CEC)}}. IEEE, \bibinfo{pages}{688--697}.
\newblock
\newblock
\shownote{doi: 10.1109/CEC.2017.7969377}.


\bibitem[Barr et~al\mbox{.}(2014)]%
        {barr2014oracle}
\bibfield{author}{\bibinfo{person}{Earl~T Barr}, \bibinfo{person}{Mark Harman}, \bibinfo{person}{Phil McMinn}, \bibinfo{person}{Muzammil Shahbaz}, {and} \bibinfo{person}{Shin Yoo}.} \bibinfo{year}{2014}\natexlab{}.
\newblock \showarticletitle{The oracle problem in software testing: A survey}.
\newblock \bibinfo{journal}{\emph{IEEE transactions on software engineering}} (\bibinfo{year}{2014}), \bibinfo{pages}{507--525}.
\newblock
\newblock
\shownote{doi: 10.1109/TSE.2014.2372785}.


\bibitem[Cao et~al\mbox{.}(2021)]%
        {cao2021work}
\bibfield{author}{\bibinfo{person}{Yumeng Cao}, \bibinfo{person}{Quinn Thibeault}, \bibinfo{person}{Aniruddh Chandratre}, \bibinfo{person}{Georgios Fainekos}, \bibinfo{person}{Giulia Pedrielli}, {and} \bibinfo{person}{Mauricio Castillo-Effen}.} \bibinfo{year}{2021}\natexlab{}.
\newblock \showarticletitle{Work-in-progress: towards assurance case evidence generation through search based testing}. In \bibinfo{booktitle}{\emph{2021 International Conference on Embedded Software (EMSOFT)}}. IEEE, \bibinfo{pages}{41--42}.
\newblock


\bibitem[Chen et~al\mbox{.}(2023)]%
        {chen2023end}
\bibfield{author}{\bibinfo{person}{Li Chen}, \bibinfo{person}{Penghao Wu}, \bibinfo{person}{Kashyap Chitta}, \bibinfo{person}{Bernhard Jaeger}, \bibinfo{person}{Andreas Geiger}, {and} \bibinfo{person}{Hongyang Li}.} \bibinfo{year}{2023}\natexlab{}.
\newblock \showarticletitle{End-to-end autonomous driving: Challenges and frontiers}.
\newblock \bibinfo{journal}{\emph{arXiv preprint arXiv:2306.16927}} (\bibinfo{year}{2023}).
\newblock


\bibitem[Ejaz et~al\mbox{.}(2013)]%
        {ejaz2013efficient}
\bibfield{author}{\bibinfo{person}{Naveed Ejaz}, \bibinfo{person}{Irfan Mehmood}, {and} \bibinfo{person}{Sung~Wook Baik}.} \bibinfo{year}{2013}\natexlab{}.
\newblock \showarticletitle{Efficient visual attention based framework for extracting key frames from videos}.
\newblock \bibinfo{journal}{\emph{Signal Processing: Image Communication}} \bibinfo{volume}{28}, \bibinfo{number}{1} (\bibinfo{year}{2013}), \bibinfo{pages}{34--44}.
\newblock


\bibitem[Feng et~al\mbox{.}(2021)]%
        {feng2021survey}
\bibfield{author}{\bibinfo{person}{Y Feng}, \bibinfo{person}{Z Xia}, \bibinfo{person}{A Guo}, {and} \bibinfo{person}{Z Chen}.} \bibinfo{year}{2021}\natexlab{}.
\newblock \showarticletitle{Survey of testing techniques of autonomous driving software}.
\newblock \bibinfo{journal}{\emph{Journal of image and Graphics}} \bibinfo{volume}{26}, \bibinfo{number}{1} (\bibinfo{year}{2021}), \bibinfo{pages}{13--27}.
\newblock


\bibitem[Fremont et~al\mbox{.}(2019)]%
        {fremont2019scenic}
\bibfield{author}{\bibinfo{person}{Daniel~J Fremont}, \bibinfo{person}{Tommaso Dreossi}, \bibinfo{person}{Shromona Ghosh}, \bibinfo{person}{Xiangyu Yue}, \bibinfo{person}{Alberto~L Sangiovanni-Vincentelli}, {and} \bibinfo{person}{Sanjit~A Seshia}.} \bibinfo{year}{2019}\natexlab{}.
\newblock \showarticletitle{Scenic: a language for scenario specification and scene generation}. In \bibinfo{booktitle}{\emph{Proceedings of the ACM SIGPLAN Conference on Programming Language Design and Implementation}}. \bibinfo{pages}{63--78}.
\newblock


\bibitem[Gambi et~al\mbox{.}(2019)]%
        {10.1145/3338906.3338942}
\bibfield{author}{\bibinfo{person}{Alessio Gambi}, \bibinfo{person}{Tri Huynh}, {and} \bibinfo{person}{Gordon Fraser}.} \bibinfo{year}{2019}\natexlab{}.
\newblock \showarticletitle{Generating Effective Test Cases for Self-Driving Cars from Police Reports}. In \bibinfo{booktitle}{\emph{Proceedings of the ACM Joint Meeting on European Software Engineering Conference and Symposium on the Foundations of Software Engineering}}.
\newblock


\bibitem[Ghahremannezhad et~al\mbox{.}(2022)]%
        {ghahremannezhad2022traffic}
\bibfield{author}{\bibinfo{person}{Hadi Ghahremannezhad}, \bibinfo{person}{Chengjun Liu}, {and} \bibinfo{person}{Hang Shi}.} \bibinfo{year}{2022}\natexlab{}.
\newblock \showarticletitle{Traffic surveillance video analytics: A concise survey}. In \bibinfo{booktitle}{\emph{Proc. 18th Int. Conf. Mach. Learn. Data Mining, New York, NY, USA}}. \bibinfo{pages}{263--291}.
\newblock


\bibitem[Guo et~al\mbox{.}(2024a)]%
        {guo2024semantic}
\bibfield{author}{\bibinfo{person}{An Guo}, \bibinfo{person}{Yang Feng}, \bibinfo{person}{Yizhen Cheng}, {and} \bibinfo{person}{Zhenyu Chen}.} \bibinfo{year}{2024}\natexlab{a}.
\newblock \showarticletitle{Semantic-guided fuzzing for virtual testing of autonomous driving systems}.
\newblock \bibinfo{journal}{\emph{Journal of Systems and Software}} (\bibinfo{year}{2024}), \bibinfo{pages}{112017}.
\newblock


\bibitem[Guo et~al\mbox{.}(2024b)]%
        {guo2024sovar}
\bibfield{author}{\bibinfo{person}{An Guo}, \bibinfo{person}{Yuan Zhou}, \bibinfo{person}{Haoxiang Tian}, \bibinfo{person}{Chunrong Fang}, \bibinfo{person}{Yunjian Sun}, \bibinfo{person}{Weisong Sun}, \bibinfo{person}{Xinyu Gao}, \bibinfo{person}{Anh~Tuan Luu}, \bibinfo{person}{Yang Liu}, {and} \bibinfo{person}{Zhenyu Chen}.} \bibinfo{year}{2024}\natexlab{b}.
\newblock \showarticletitle{SoVAR: Building Generalizable Scenarios from Accident Reports for Autonomous Driving Testing}.
\newblock \bibinfo{journal}{\emph{arXiv preprint arXiv:2409.08081}} (\bibinfo{year}{2024}).
\newblock


\bibitem[Hekmatnejad et~al\mbox{.}(2020)]%
        {hekmatnejad2020search}
\bibfield{author}{\bibinfo{person}{Mohammad Hekmatnejad}, \bibinfo{person}{Bardh Hoxha}, {and} \bibinfo{person}{Georgios Fainekos}.} \bibinfo{year}{2020}\natexlab{}.
\newblock \showarticletitle{Search-based test-case generation by monitoring responsibility safety rules}. In \bibinfo{booktitle}{\emph{IEEE International Conference on Intelligent Transportation Systems (ITSC)}}.
\newblock
\newblock
\shownote{doi: 10.1109/ITSC45102.2020.9294489}.


\bibitem[Hungar(2018)]%
        {hungar2018scenario}
\bibfield{author}{\bibinfo{person}{Hardi Hungar}.} \bibinfo{year}{2018}\natexlab{}.
\newblock \showarticletitle{Scenario-based validation of automated driving systems}. In \bibinfo{booktitle}{\emph{International Symposium on Leveraging Applications of Formal Methods}}. Springer, \bibinfo{pages}{449--460}.
\newblock


\bibitem[Ishak and Abu~Bakar(2014)]%
        {ishak2014developing}
\bibfield{author}{\bibinfo{person}{Noriah~Mohd Ishak} {and} \bibinfo{person}{Abu~Yazid Abu~Bakar}.} \bibinfo{year}{2014}\natexlab{}.
\newblock \showarticletitle{Developing Sampling Frame for Case Study: Challenges and Conditions.}
\newblock \bibinfo{journal}{\emph{World journal of education}} \bibinfo{volume}{4}, \bibinfo{number}{3} (\bibinfo{year}{2014}), \bibinfo{pages}{29--35}.
\newblock


\bibitem[Jullien et~al\mbox{.}(2009)]%
        {jullien2009openscenario}
\bibfield{author}{\bibinfo{person}{Jean-Michel Jullien}, \bibinfo{person}{Christian Martel}, \bibinfo{person}{Laurence Vignollet}, {and} \bibinfo{person}{Maia Wentland}.} \bibinfo{year}{2009}\natexlab{}.
\newblock \showarticletitle{OpenScenario: a flexible integrated environment to develop educational activities based on pedagogical scenarios}. In \bibinfo{booktitle}{\emph{2009 Ninth IEEE International Conference on Advanced Learning Technologies}}. IEEE, \bibinfo{pages}{509--513}.
\newblock


\bibitem[Li et~al\mbox{.}(2020)]%
        {li2020av}
\bibfield{author}{\bibinfo{person}{Guanpeng Li}, \bibinfo{person}{Yiran Li}, \bibinfo{person}{Saurabh Jha}, \bibinfo{person}{Timothy Tsai}, \bibinfo{person}{Michael Sullivan}, \bibinfo{person}{Siva Kumar~Sastry Hari}, \bibinfo{person}{Zbigniew Kalbarczyk}, {and} \bibinfo{person}{Ravishankar Iyer}.} \bibinfo{year}{2020}\natexlab{}.
\newblock \showarticletitle{Av-fuzzer: Finding safety violations in autonomous driving systems}. In \bibinfo{booktitle}{\emph{Proceedings of IEEE International Symposium on Software Reliability Engineering (ISSRE)}}. \bibinfo{pages}{25--36}.
\newblock
\newblock
\shownote{doi: 10.1109/ISSRE5003.2020.00012}.


\bibitem[Lou et~al\mbox{.}(2022)]%
        {lou2022testing}
\bibfield{author}{\bibinfo{person}{Guannan Lou}, \bibinfo{person}{Yao Deng}, \bibinfo{person}{Xi Zheng}, \bibinfo{person}{Mengshi Zhang}, {and} \bibinfo{person}{Tianyi Zhang}.} \bibinfo{year}{2022}\natexlab{}.
\newblock \showarticletitle{Testing of autonomous driving systems: where are we and where should we go?}. In \bibinfo{booktitle}{\emph{Proceedings of the 30th ACM Joint European Software Engineering Conference and Symposium on the Foundations of Software Engineering}}. \bibinfo{pages}{31--43}.
\newblock


\bibitem[Manning et~al\mbox{.}(2014)]%
        {manning2014stanford}
\bibfield{author}{\bibinfo{person}{Christopher~D Manning}, \bibinfo{person}{Mihai Surdeanu}, \bibinfo{person}{John Bauer}, \bibinfo{person}{Jenny~Rose Finkel}, \bibinfo{person}{Steven Bethard}, {and} \bibinfo{person}{David McClosky}.} \bibinfo{year}{2014}\natexlab{}.
\newblock \showarticletitle{The Stanford CoreNLP natural language processing toolkit}. In \bibinfo{booktitle}{\emph{Proceedings of 52nd annual meeting of the association for computational linguistics: system demonstrations}}. \bibinfo{pages}{55--60}.
\newblock


\bibitem[Narayanan et~al\mbox{.}(2019)]%
        {narayanan2019dynamic}
\bibfield{author}{\bibinfo{person}{Athma Narayanan}, \bibinfo{person}{Isht Dwivedi}, {and} \bibinfo{person}{Behzad Dariush}.} \bibinfo{year}{2019}\natexlab{}.
\newblock \showarticletitle{Dynamic Traffic Scene Classification with Space-Time Coherence}.
\newblock \bibinfo{journal}{\emph{arXiv preprint arXiv:1905.12708}} (\bibinfo{year}{2019}).
\newblock


\bibitem[Nejati(2019)]%
        {nejati2019testing}
\bibfield{author}{\bibinfo{person}{Shiva Nejati}.} \bibinfo{year}{2019}\natexlab{}.
\newblock \showarticletitle{Testing cyber-physical systems via evolutionary algorithms and machine learning}. In \bibinfo{booktitle}{\emph{2019 IEEE/ACM 12th International Workshop on Search-Based Software Testing (SBST)}}. IEEE, \bibinfo{pages}{1--1}.
\newblock
\newblock
\shownote{doi: 10.1109/SBST.2019.00008}.


\bibitem[Neumann and Deml(2011)]%
        {neumann2011two}
\bibfield{author}{\bibinfo{person}{Hendrik Neumann} {and} \bibinfo{person}{Barbara Deml}.} \bibinfo{year}{2011}\natexlab{}.
\newblock \showarticletitle{The two-point visual control model of steering-new empirical evidence}. In \bibinfo{booktitle}{\emph{Digital Human Modeling: Third International Conference, ICDHM 2011, Held as Part of HCI International 2011, Orlando, FL, USA July 9-14, 2011. Proceedings 3}}. Springer, \bibinfo{pages}{493--502}.
\newblock


\bibitem[Plyer et~al\mbox{.}(2016)]%
        {plyer2016massively}
\bibfield{author}{\bibinfo{person}{Aur{\'e}lien Plyer}, \bibinfo{person}{Guy Le~Besnerais}, {and} \bibinfo{person}{Fr{\'e}d{\'e}ric Champagnat}.} \bibinfo{year}{2016}\natexlab{}.
\newblock \showarticletitle{Massively parallel Lucas Kanade optical flow for real-time video processing applications}.
\newblock \bibinfo{journal}{\emph{Journal of Real-Time Image Processing}}  \bibinfo{volume}{11} (\bibinfo{year}{2016}), \bibinfo{pages}{713--730}.
\newblock


\bibitem[Ramanishka et~al\mbox{.}(2018)]%
        {ramanishka2018CVPR}
\bibfield{author}{\bibinfo{person}{Vasili Ramanishka}, \bibinfo{person}{Yi-Ting Chen}, \bibinfo{person}{Teruhisa Misu}, {and} \bibinfo{person}{Kate Saenko}.} \bibinfo{year}{2018}\natexlab{}.
\newblock \showarticletitle{Toward Driving Scene Understanding: A Dataset for Learning Driver Behavior and Causal Reasoning}. In \bibinfo{booktitle}{\emph{Conference on Computer Vision and Pattern Recognition (CVPR)}}.
\newblock


\bibitem[Romera-Paredes et~al\mbox{.}(2024)]%
        {romera2024mathematical}
\bibfield{author}{\bibinfo{person}{Bernardino Romera-Paredes}, \bibinfo{person}{Mohammadamin Barekatain}, \bibinfo{person}{Alexander Novikov}, \bibinfo{person}{Matej Balog}, \bibinfo{person}{M~Pawan Kumar}, \bibinfo{person}{Emilien Dupont}, \bibinfo{person}{Francisco~JR Ruiz}, \bibinfo{person}{Jordan~S Ellenberg}, \bibinfo{person}{Pengming Wang}, \bibinfo{person}{Omar Fawzi}, {et~al\mbox{.}}} \bibinfo{year}{2024}\natexlab{}.
\newblock \showarticletitle{Mathematical discoveries from program search with large language models}.
\newblock \bibinfo{journal}{\emph{Nature}} \bibinfo{volume}{625}, \bibinfo{number}{7995} (\bibinfo{year}{2024}), \bibinfo{pages}{468--475}.
\newblock


\bibitem[Salvucci and Gray(2004)]%
        {salvucci2004two}
\bibfield{author}{\bibinfo{person}{Dario~D Salvucci} {and} \bibinfo{person}{Rob Gray}.} \bibinfo{year}{2004}\natexlab{}.
\newblock \showarticletitle{A two-point visual control model of steering}.
\newblock \bibinfo{journal}{\emph{Perception}} \bibinfo{volume}{33}, \bibinfo{number}{10} (\bibinfo{year}{2004}), \bibinfo{pages}{1233--1248}.
\newblock


\bibitem[Sarker et~al\mbox{.}(2020)]%
        {sarker2020screw}
\bibfield{author}{\bibinfo{person}{Anik Sarker}, \bibinfo{person}{Anirban Sinha}, {and} \bibinfo{person}{Nilanjan Chakraborty}.} \bibinfo{year}{2020}\natexlab{}.
\newblock \showarticletitle{On screw linear interpolation for point-to-point path planning}. In \bibinfo{booktitle}{\emph{2020 IEEE/RSJ International Conference on Intelligent Robots and Systems (IROS)}}. IEEE, \bibinfo{pages}{9480--9487}.
\newblock


\bibitem[Shih(2013)]%
        {shih2013novel}
\bibfield{author}{\bibinfo{person}{Huang-Chia Shih}.} \bibinfo{year}{2013}\natexlab{}.
\newblock \showarticletitle{A novel attention-based key-frame determination method}.
\newblock \bibinfo{journal}{\emph{IEEE Transactions on Broadcasting}} \bibinfo{volume}{59}, \bibinfo{number}{3} (\bibinfo{year}{2013}), \bibinfo{pages}{556--562}.
\newblock


\bibitem[Shin et~al\mbox{.}(2018)]%
        {shin2018test}
\bibfield{author}{\bibinfo{person}{Seung~Yeob Shin}, \bibinfo{person}{Shiva Nejati}, \bibinfo{person}{Mehrdad Sabetzadeh}, \bibinfo{person}{Lionel~C Briand}, {and} \bibinfo{person}{Frank Zimmer}.} \bibinfo{year}{2018}\natexlab{}.
\newblock \showarticletitle{Test case prioritization for acceptance testing of cyber physical systems: a multi-objective search-based approach}. In \bibinfo{booktitle}{\emph{Proceedings of the acm sigsoft international symposium on software testing and analysis}}. \bibinfo{pages}{49--60}.
\newblock
\newblock
\shownote{doi: 10.1145/3213846.3213852}.


\bibitem[Singh(2015)]%
        {singh2015critical}
\bibfield{author}{\bibinfo{person}{Santokh Singh}.} \bibinfo{year}{2015}\natexlab{}.
\newblock \bibinfo{booktitle}{\emph{Critical reasons for crashes investigated in the national motor vehicle crash causation survey}}.
\newblock \bibinfo{type}{{T}echnical {R}eport}.
\newblock


\bibitem[Singh and Saini(2021)]%
        {singh2021autonomous}
\bibfield{author}{\bibinfo{person}{Sehajbir Singh} {and} \bibinfo{person}{Baljit~Singh Saini}.} \bibinfo{year}{2021}\natexlab{}.
\newblock \showarticletitle{Autonomous cars: Recent developments, challenges, and possible solutions}. In \bibinfo{booktitle}{\emph{IOP conference series: Materials science and engineering}}, Vol.~\bibinfo{volume}{1022}. IOP Publishing, \bibinfo{pages}{012028}.
\newblock


\bibitem[Song et~al\mbox{.}(2021)]%
        {song2021dimension}
\bibfield{author}{\bibinfo{person}{Shiming Song}, \bibinfo{person}{Pengjun Wang}, \bibinfo{person}{Ali~Asghar Heidari}, \bibinfo{person}{Mingjing Wang}, \bibinfo{person}{Xuehua Zhao}, \bibinfo{person}{Huiling Chen}, \bibinfo{person}{Wenming He}, {and} \bibinfo{person}{Suling Xu}.} \bibinfo{year}{2021}\natexlab{}.
\newblock \showarticletitle{Dimension decided Harris hawks optimization with Gaussian mutation: Balance analysis and diversity patterns}.
\newblock \bibinfo{journal}{\emph{Knowledge-Based Systems}}  \bibinfo{volume}{215} (\bibinfo{year}{2021}), \bibinfo{pages}{106425}.
\newblock


\bibitem[Tang et~al\mbox{.}(2024)]%
        {tang2024legend}
\bibfield{author}{\bibinfo{person}{Shuncheng Tang}, \bibinfo{person}{Zhenya Zhang}, \bibinfo{person}{Jixiang Zhou}, \bibinfo{person}{Lei Lei}, \bibinfo{person}{Yuan Zhou}, {and} \bibinfo{person}{Yinxing Xue}.} \bibinfo{year}{2024}\natexlab{}.
\newblock \showarticletitle{LeGEND: A Top-Down Approach to Scenario Generation of Autonomous Driving Systems Assisted by Large Language Models}.
\newblock \bibinfo{journal}{\emph{arXiv preprint arXiv:2409.10066}} (\bibinfo{year}{2024}).
\newblock


\bibitem[Tian et~al\mbox{.}(2022a)]%
        {tian2022mosat}
\bibfield{author}{\bibinfo{person}{Haoxiang Tian}, \bibinfo{person}{Yan Jiang}, \bibinfo{person}{Guoquan Wu}, \bibinfo{person}{Jiren Yan}, \bibinfo{person}{Jun Wei}, \bibinfo{person}{Wei Chen}, \bibinfo{person}{Shuo Li}, {and} \bibinfo{person}{Dan Ye}.} \bibinfo{year}{2022}\natexlab{a}.
\newblock \showarticletitle{MOSAT: finding safety violations of autonomous driving systems using multi-objective genetic algorithm}. In \bibinfo{booktitle}{\emph{Proceedings of the 30th ACM Joint European Software Engineering Conference and Symposium on the Foundations of Software Engineering}}. \bibinfo{pages}{94--106}.
\newblock


\bibitem[Tian et~al\mbox{.}(2022b)]%
        {tian2022generating}
\bibfield{author}{\bibinfo{person}{Haoxiang Tian}, \bibinfo{person}{Guoquan Wu}, \bibinfo{person}{Jiren Yan}, \bibinfo{person}{Yan Jiang}, \bibinfo{person}{Jun Wei}, \bibinfo{person}{Wei Chen}, \bibinfo{person}{Shuo Li}, {and} \bibinfo{person}{Dan Ye}.} \bibinfo{year}{2022}\natexlab{b}.
\newblock \showarticletitle{Generating critical test scenarios for autonomous driving systems via influential behavior patterns}. In \bibinfo{booktitle}{\emph{Proceedings of the 37th IEEE/ACM International Conference on Automated Software Engineering}}. \bibinfo{pages}{1--12}.
\newblock


\bibitem[Wang et~al\mbox{.}(2016)]%
        {10.1145/2884781.2884880}
\bibfield{author}{\bibinfo{person}{Shuai Wang}, \bibinfo{person}{Shaukat Ali}, \bibinfo{person}{Tao Yue}, \bibinfo{person}{Yan Li}, {and} \bibinfo{person}{Marius Liaaen}.} \bibinfo{year}{2016}\natexlab{}.
\newblock \showarticletitle{A Practical Guide to Select Quality Indicators for Assessing Pareto-Based Search Algorithms in Search-Based Software Engineering}. In \bibinfo{booktitle}{\emph{Proceedings of International Conference on Software Engineering (ICSE)}}. \bibinfo{pages}{631--642}.
\newblock
\newblock
\shownote{doi: 10.1145/2884781.2884880}.


\bibitem[Wang et~al\mbox{.}(2020)]%
        {wang2020abnormal}
\bibfield{author}{\bibinfo{person}{Tian Wang}, \bibinfo{person}{Meina Qiao}, \bibinfo{person}{Aichun Zhu}, \bibinfo{person}{Guangcun Shan}, {and} \bibinfo{person}{Hichem Snoussi}.} \bibinfo{year}{2020}\natexlab{}.
\newblock \showarticletitle{Abnormal event detection via the analysis of multi-frame optical flow information}.
\newblock \bibinfo{journal}{\emph{Frontiers of Computer Science}}  \bibinfo{volume}{14} (\bibinfo{year}{2020}), \bibinfo{pages}{304--313}.
\newblock


\bibitem[Zhang and Cai(2023)]%
        {zhang2023building}
\bibfield{author}{\bibinfo{person}{Xudong Zhang} {and} \bibinfo{person}{Yan Cai}.} \bibinfo{year}{2023}\natexlab{}.
\newblock \showarticletitle{Building critical testing scenarios for autonomous driving from real accidents}. In \bibinfo{booktitle}{\emph{Proceedings of the 32nd ACM SIGSOFT International Symposium on Software Testing and Analysis}}. \bibinfo{pages}{462--474}.
\newblock


\bibitem[Zhou et~al\mbox{.}(2023)]%
        {zhou2023specification}
\bibfield{author}{\bibinfo{person}{Yuan Zhou}, \bibinfo{person}{Yang Sun}, \bibinfo{person}{Yun Tang}, \bibinfo{person}{Yuqi Chen}, \bibinfo{person}{Jun Sun}, \bibinfo{person}{Christopher~M Poskitt}, \bibinfo{person}{Yang Liu}, {and} \bibinfo{person}{Zijiang Yang}.} \bibinfo{year}{2023}\natexlab{}.
\newblock \showarticletitle{Specification-based autonomous driving system testing}.
\newblock \bibinfo{journal}{\emph{IEEE Transactions on Software Engineering}} (\bibinfo{year}{2023}).
\newblock


\bibitem[Zipfl et~al\mbox{.}(2023)]%
        {zipfl2023comprehensive}
\bibfield{author}{\bibinfo{person}{Maximilian Zipfl}, \bibinfo{person}{Nina Koch}, {and} \bibinfo{person}{J~Marius Z{\"o}llner}.} \bibinfo{year}{2023}\natexlab{}.
\newblock \showarticletitle{A comprehensive review on ontologies for scenario-based testing in the context of autonomous driving}. In \bibinfo{booktitle}{\emph{2023 IEEE Intelligent Vehicles Symposium (IV)}}. IEEE, \bibinfo{pages}{1--7}.
\newblock


\end{thebibliography}


\end{document}